\documentclass[apj]{emulateapj}
\newcommand{\lam}{$\lambda$}
\newcommand\rmxaa{\ref@jnl{Rev. Mexicana Astron. Astrofis.}}%
\usepackage{epsfig}
\shorttitle{Independent Emission and Absorption Abundances}
\shortauthors{Williams et al.}
\slugcomment{The Astrophysical Journal, in press, vol. 677, No. 2 (20
April 2008)}

\begin{document}

\title{Independent Emission and Absorption Abundances for Planetary
Nebulae\footnote{Based on observations with the NASA/ESA Hubble Space
Telescope obtained at the Space Telescope Science Institute, which is
operated by AURA, Inc. under NASA contract NAS5-26555.}}

\author{Robert Williams\altaffilmark{2}, Edward
B. Jenkins\altaffilmark{3}, Jack A. Baldwin\altaffilmark{4}, Yong
Zhang\altaffilmark{5}, Brian Sharpee\altaffilmark{6}, Eric
Pellegrini\altaffilmark{4}, and Mark Phillips\altaffilmark{7}}

\altaffiltext{2}{Space Telescope Science Institute, 3700 San Martin
Drive, Baltimore, MD 21218.}

\altaffiltext{3}{Princeton University Observatory, Princeton, NJ
08544.}

\altaffiltext{4}{Department of Physics \& Astronomy, Michigan State
University, East Lansing, MI 48824.}

\altaffiltext{5}{Department of Physics, University of Hong Kong, Hong
Kong, China.}

\altaffiltext{6}{Molecular Physics Laboratory, SRI International, 333
Ravenswood Avenue, Menlo Park, CA 94025.}

\altaffiltext{7}{Las Campanas Observatory, Carnegie Observatories,
Casilla 601, La Serena, Chile.}

\begin{abstract}

Emission-line abundances have been uncertain for more than a decade
due to unexplained discrepancies in the relative intensities of the
forbidden lines and weak permitted recombination lines in planetary
nebulae (PNe) and H II regions.  The observed intensities of forbidden
and recombination lines originating from the same parent ion differ
from their theoretical values by factors of more than an order of
magnitude in some of these nebulae.  In this study we observe UV
resonance line absorption in the central stars of PNe produced by the
nebular gas, and from the same ions that emit optical forbidden lines.
We then compare the derived absorption column densities with the
emission measures determined from ground-based observations of the
nebular forbidden lines.  We find for our sample of PNe that the
collisionally excited forbidden lines yield column densities that are
in basic agreement with the column densities derived for the same ions
from the UV absorption lines.  A similar comparison involving
recombination line column densities produces poorer agreement,
although near the limits of the formal uncertainties of the analyses.
An additional sample of objects with larger abundance discrepancy
factors will need to be studied before a stronger statement can be
made that recombination line abundances are not correct.
\end{abstract}


\section{Introduction}

The analysis of emission line intensities has been used to determine
nebular abundances for a wide range of objects.  Standard procedures
have been developed in which the collisionally excited forbidden lines
and high-level permitted recombination lines of ions are used to
determine abundances \citep{DS03,OF06}.  For most objects heavy
element abundances are derived from the forbidden lines because of
their greater strengths compared to the fainter recombination lines,
which are frequently only marginally stronger than the continuum
intensity.  For some of the higher surface brightness gaseous nebulae
both types of lines have been used to determine the heavy element CNO
abundances, and they have produced discrepant results of more than an
order of magnitude in some objects.

Each of the two types of lines have certain advantages for abundance
determinations.  The forbidden lines are strong, so they are detected
from many more ions than the recombination lines.  The collisional
excitation of low-lying levels dominates other competing population
processes such as fluorescence excitation, charge exchange, and
dielectronic recombination.  Furthermore, collision strengths coupling
most of the lower bound levels of ions are known to better than 30\%
accuracy.  The largest uncertainty in using forbidden line intensities
for abundances is their sensitivity to kinetic temperature that
results from excitation by electron impact.

\begin{deluxetable*}{lp{3.5in}p{2in}}
\tablecaption{Ions with UV Resonance and Optical Forbidden Lines
  \label{tab1}}
\tablewidth{6.5in}
\tablehead{
\colhead{Ion} &
\multicolumn{1}{c}{UV Resonance Transition $\lambda$(\AA)\tablenotemark{a}} &
\multicolumn{1}{c}{Optical Forbidden Line  $\lambda$(\AA)}
}
\startdata
\ion{C}{1} & 1277.5*, 1329.6*, 1561.4*, 1657.0* & 4622,
8727, 9850 \\
\ion{P}{2} & 1152.8, 1154.0*, 1301.9, 1310.7*, 1542.3*,
1532.5 & 4669, 7876 \\
\ion{S}{1} & 1270.8, 1277.2, 1295.7, 1316.5, 1425.0, 1433.3*,
1474.0, 1483.0*, 1807.3, 1820.3* & 4589, 7725 \\
\ion{Fe}{2} & 1260.5, 1608.5, 1621.7* & 4244, 4359, 5159, 8617 \\
\ion{Ni}{2} & 1317.2, 1370.1, 1454.8, 1709.6, 1741.5 & 6667, 7378 \\
\ion{N}{1} & 1199.5, 1200.2, 1200.7 & 3467, 5198, 5200, 10398 \\
\ion{O}{1} & 1302.2, 1304.9*, 1306.0** & 5577, 6300, 6364 \\ 
\ion{S}{2} & 1250.6, 1253.8, 1259.5 & 4069, 4076, 6716, 6731, 10320 \\
\ion{S}{3} & 1190.2, 1194.1*, 1201.7* & 3722, 6312,
9069, 9531 \\
\enddata
\tablenotetext{a}{Single and double asterisks indicate transitions
arising from the first and second fine-structure excited level of
ground-state term, respectively.}
\end{deluxetable*}

Direct electron recapture populates the higher levels of ions and this
process has relatively small cross sections.  Thus, recombination
lines tend not to be strong except for H and He by virtue of their
dominant abundances, but they are observable in nebulae from ions of
CNO and Ne.  They have the advantage that recombination line intensity
ratios are insensitive to temperature and density, and the relevant
cross sections are believed to be known reasonably well.  Because
recombination cross sections are small, however, other excitation
processes compete with electron recapture in populating the higher
levels from which these lines are observed.  Thus, there can be
greater uncertainty in the excitation processes that are responsible
for specific high level permitted lines.

\begin{deluxetable*}{p{2.6in}cccc}
\tablecaption{Journal of Observations for HST UV Spectroscopy
  \label{tab2}}
\tablewidth{6.5in}
\tablehead{
\multicolumn{1}{c}{Object} &
\colhead{He2-138} &
\colhead{NGC 246} &
\colhead{NGC 6543} &
\colhead{Tc 1}
}
\startdata 
Central star (V) & 10.9 & 11.9 & 11.1 & 11.4 \\
Shell surface brightness, S(H$\beta$) \\ \hspace{0.25in}(erg cm$^{-2}$ s$^{-1}$
arcsec$^2$) & 5.1$\times 10^{-13}$ & 6.2$\times 10^{-16}$ & 8.2$\times
10^{-13}$ & 3.0$\times 10^{-13}$ \\
Diameter of Central Nebular \\ \hspace{0.25in}Emission (arcsec) & 7 & 245 & 20 & 10 \\
Radial velocity, heliocentric (km s$^{-1}$) & -47 & -46 & -66 & -83 \\
Exposure times (sec): & \\
\multicolumn{1}{r}{1150--1330\AA} & 7$\times$2114 & 1967 & 2$\times$2150
& 2$\times$2072 \\
 & & & & 8$\times$1365 \\
\multicolumn{1}{r}{1316--1518\AA} & \nodata & 2$\times$1368 & 2440 &
\nodata \\
\multicolumn{1}{r}{1495--1688\AA} & 7$\times$2116 & \nodata &
3$\times$2460 & 2077 \\
 & & & & 4$\times$1367 \\
\enddata
\end{deluxetable*}

Electron temperatures and densities are determined directly from the
relative intensities of forbidden lines originating on different
levels of the same ion with the result that emission spectra have been
a major source of our knowledge of element abundances of every type of
emission-line object.  The relatively high surface brightnesses of
planetary nebulae and a few of the brighter H II regions enable the
recombination lines of CNO to be observed, and in the past decade ion
abundances have been determined for a number of PNe using both the
forbidden lines (FL) and the recombination lines (RL) from the same
ions.  Surprisingly, the two types of lines have not yielded the same
abundances.  The differences between the RL and FL abundances vary
from object to object and span the range from 15 percent, i.e.,
relatively good agreement, to factors of 50 and more
\citep{Ts04,RG05,Li06,GE07}, with the recombination line intensities
always being stronger than predicted relative to the forbidden lines
and therefore indicative of higher abundances.  These discrepancies
have been the subject of many studies which have given rise to a large
literature on the subject, but they are still not understood.  Until
they are resolved some doubt is cast upon the normal methods by which
collisionally excited forbidden lines are used to derive element
abundances.  The differences cannot be due to incorrect atomic data
since this would cause the magnitude of the discrepancies to be
roughly the same for all objects. Current resolutions to the
discrepancies problem have focused on temperature fluctuations in the
nebulae \citep{Pe67,Pe04} and dense inclusions that are hydrogen
deficient \citep{Li00}.
 
A completely independent, alternative method of obtaining abundances
for nebular gas does exist and can be used as an independent check of
emission-line abundances.  It involves observing the absorption lines
produced by the foreground nebular gas in the spectrum of an embedded
or background star to determine column densities.  Most of the
absorption lines occur in the ultraviolet because low density gas
occupies the ground state and the resonance lines of the most
cosmically abundant ions fall in the UV.  Thus, a space telescope with
a high resolution spectrograph is required to study these absorption
lines with sufficient resolution to yield reliable column densities.

\begin{figure}[t]
\epsscale{}
\plotone{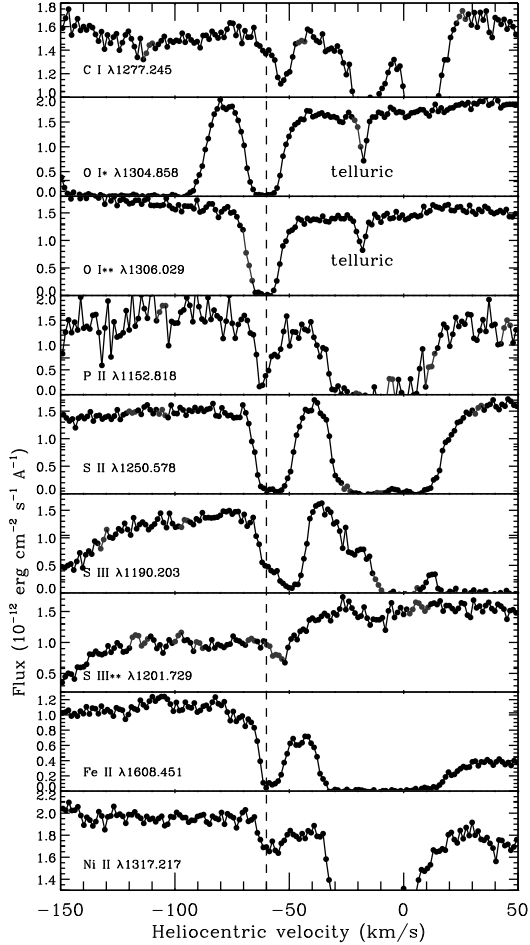}
\caption{A montage of UV absorption line profiles from HST/STIS
spectra of the central star of He2-138.  The foreground ISM absorption
is centered around velocity = -10 km s$^{-1}$, and the PN shell
absorption is centered around velocity = -60 km s$^{-1}$, as indicated
by the vertical dashed line. \label{fig1}
}
\end{figure}

\begin{figure}[th]
%\epsscale{0.5}
\plotone{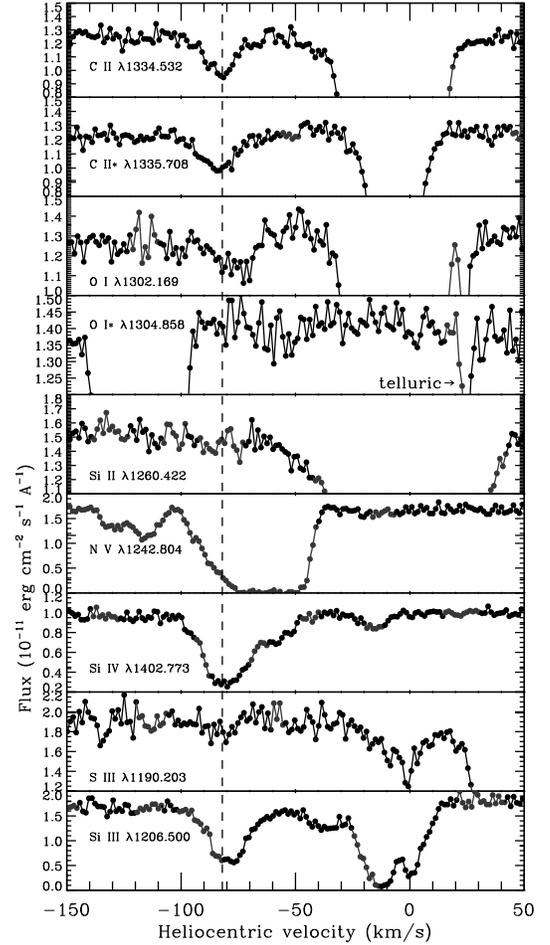}
\caption{A montage of UV absorption line profiles from the star of NGC
246 from the HST/STIS data.  The ISM absorption is centered near
velocity = 0, and the PN shell absorption is centered around velocity
= -80 km s$^{-1}$, as indicated by the vertical dashed line.  The
broad \ion{N}{5} absorption trough is indicative of an outflowing wind
from the central star, which has a heliocentric velocity of v* = -46
km s$^{-1}$. \label{fig2} 
}
\end{figure}

\begin{figure}[t]
%\epsscale{0.5}
\plotone{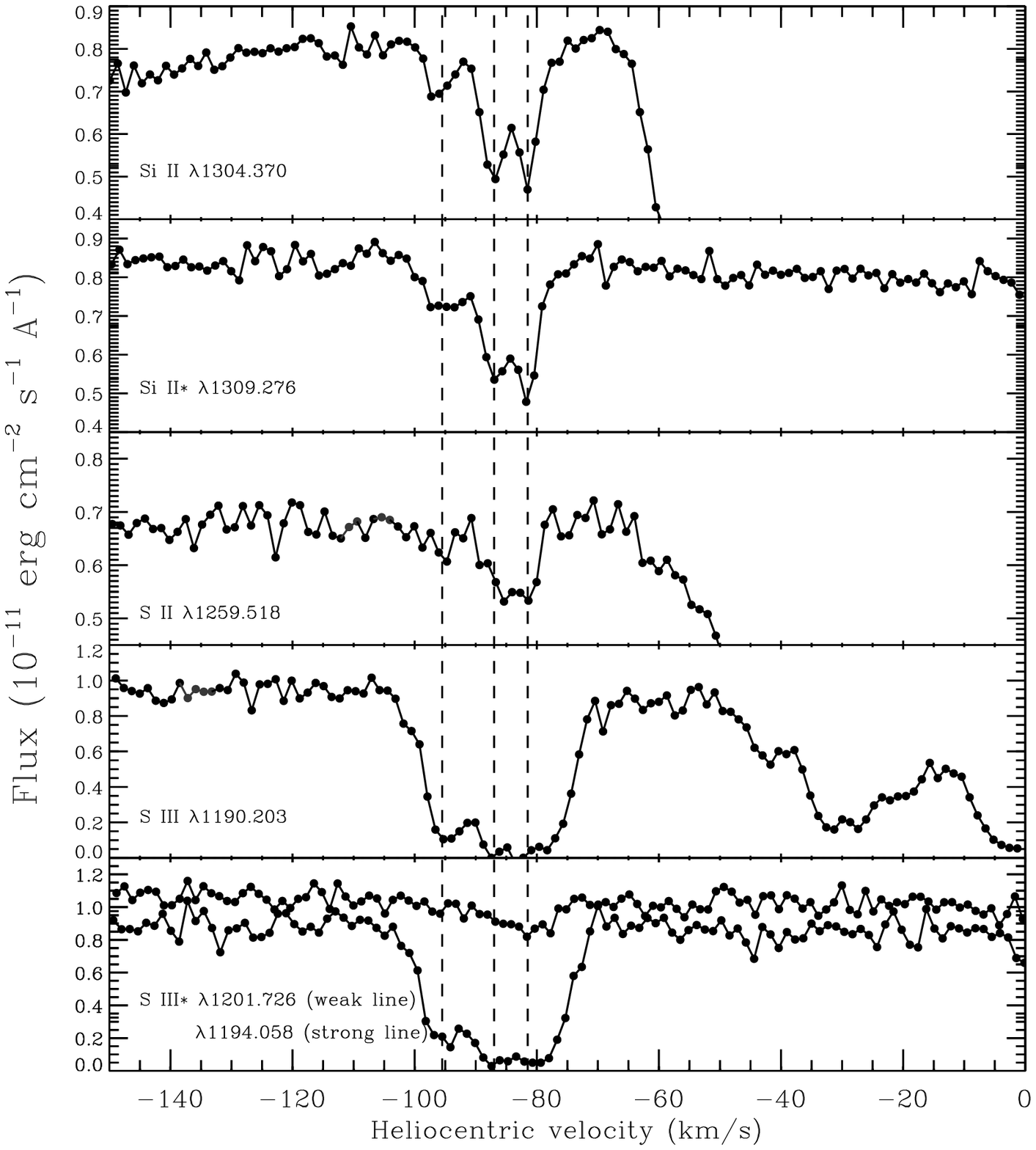}
\caption{Same as Figure~\ref{fig1} except for the central star
NGC~6543.  The three vertical dashed lines represent distinct
absorption components from the nebular shell. \label{fig3} 
}
\end{figure}

\citet{PMP84} and \citet{PPM86} made the first attempts to obtain ion
abundances in planetary nebulae by this method, using the IUE high
resolution spectrograph (R=15,000) to measure absorption line
equivalent widths from which column densities were determined.  For
the two PNe whose central stars were bright enough to be studied with
IUE, the relatively few unsaturated lines for which they were able to
obtain column densities belonged to ions that do not have detectable
emission lines in the optical.  Hence, although they did determine
relative abundances from absorption line data, they were unable to
make a direct comparison between independently derived absorption and
emission line abundances for the same ions.

The high resolution spectrographs of HST increase the number of
nebulae for which reliable column densities can be obtained, and offer
the possibility of resolving the question of the abundance
discrepancies.  The more abundant heavy element ions having resonance
lines in the UV between $\lambda\lambda$1150-1800\AA\ accessible with
the STIS spectrograph and also having detectable forbidden lines at
optical wavelengths are listed in Table~\ref{tab1}.  The abundances of
these ions can be determined independently from ground and space
telescopes by completely different methods and then compared, albeit
having separate sight lines and path lengths, e.g., the line of sight
to the star does not probe the rear part of the nebula.

Specifically, the column densities of individual ions can be found
from the UV absorption spectra, while emission measures are derived
from the forbidden and permitted nebular emission lines.  For each ion
the column density and emission measure differ only by the
multiplicative factor of the electron density in the emission measure.
Since standard nebular diagnostics provide a direct determination of
the density appropriate for each ion depending upon its ionization
level, a direct comparison can be made between the absorption column
density of the ion and its emission measure as derived from the
different emission lines.  This procedure should demonstrate which
emission lines, forbidden or permitted, yield abundances most
consistent with those from the UV absorption lines.

An initial study of abundances determined from UV absorption lines in
the central star vs. those found from nebular emission line
intensities for the PN IC 418 was attempted by \citet{Wi03}.  For the
four ions S$^{+2}$, S$^+$, Ni$^+$, and Fe$^+$, and netural oxygen
O$^o$, for which relative abundances could be determined independently
from both methods, rough agreement was found.  However, the
uncertainties were too large for meaningful conclusions to be drawn.

We report here on an observing program which attempts to resolve the
discrepancies between the forbidden and permitted emission line
intensities by making a UV absorption line analysis that independently
serves to validate emission line results.  We have obtained high
resolution UV spectra of four PNe central stars with HST/STIS, and
visible spectra of three of the associated nebular shells from Las
Campanas and KPNO.  The UV observations and absorption analysis are
described in \S\ref{sec3}, and the optical emission spectra and
analysis are presented in \S\ref{sec4}.  The relative column densities
from the two methods are compared and interpreted in \S\ref{sec5}.

\section{Object Sample and Observing Program} \label{sec1}

Column densities determined from absorption lines are most reliable
when the lines are well-resolved and have ample signal-to-noise to
define the continuum, thus brighter central stars are advantageous.
Absorption lines originating in nebular gas are frequently seriously
blended with and obliterated by stronger absorption from the same
transitions caused by intervening ISM gas along the same line of
sight.  Unambiguous measurement of absorption from the nebular gas
therefore requires a nebular radial velocity differing by at least 50
km s$^{-1}$ from that of the Local Standard of Rest to shift the
nebular absorption out of the corresponding stronger ISM component.
Optimal candidates for emission study are preferentially high surface
brightness objects, thus favoring PNe over the lower surface
brightness H II regions.  It would be advantageous to include in our
sample some PNe for which the largest FL and RL abundance
discrepancies have been determined, however the few PNe that have been
established to have differences of more than a factor of ten either
have (a) central stars that are too faint in the UV, (b) very low
surface brightnesses, or (c) radial velocities that are not
sufficiently different from the LSR to avoid confusion between the
nebular shell and ISM absorption lines.

\begin{figure}[t]
%\epsscale{0.5}
\plotone{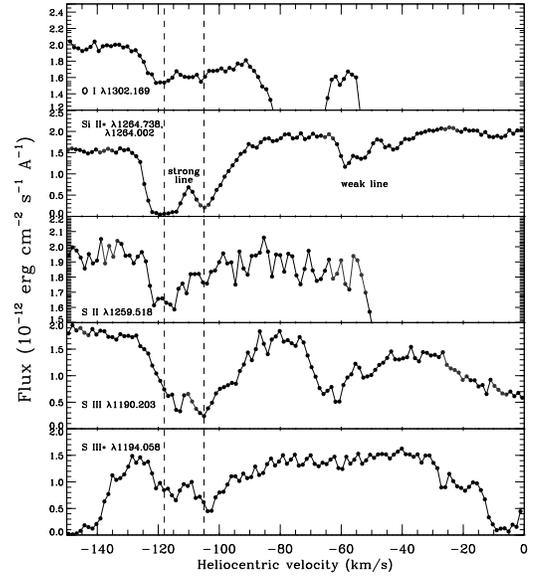}
\caption{Same as Figure~\ref{fig1}, except for the central star of Tc
  1. \label{fig4} }
\end{figure}

The sample of known PNe satisfying the optimal criteria for study is
given in \citet{Wi03}, and is not large.  We identified four PNe that
satisfy these criteria and which seem well suited for a combined
UV-visible study, viz., He2-138, NGC~246, NGC~6543, \& Tc 1, whose
central stars are sufficiently bright that UV observations with HST at
high spectral resolution would produce acceptable spectra in
reasonable exposure times.  We acquired spectra of the central stars
of these PNe in the UV with HST/STIS and then subsequently observed
the nebular shells along adjacent sight lines in the visible with
ground-based telescopes to obtain line intensities for the
emission-line analysis.

\section{Absorption-Line Analysis} \label{sec3}

\subsection{STIS Observations} \label{sec3.1}

HST/STIS was used to obtain spectra of the central stars of He2-138,
NGC~246, NGC~6543, and Tc 1 in the high resolution mode, i.e., grating
E140H with a resolution of 3 km s$^{-1}$, in three separate settings
that covered the wavelength region 1150--1690 \AA.  Exposure times
that produced a continuum signal-to-noise level of S/N$\approx$15 over
the entire wavelength regime for each grating setting were adopted.
The observations were made in 2005 (Cycle 12), and the relevant
properties of our targets and the journal of the STIS observations is
given in Table~\ref{tab2}.  Regrettably, STIS failed and became
inoperative before our observing program could be completed, thus we
did not succeed in executing all of our planned observations.  Only
partial data exist for each of the central stars.  The spectra were
reduced using the most recent version of CALSTIS procedures and
algorithms \citep{Li98}, and a montage of resonance line profiles from
the final reduced spectra that include ions for which we also
subsequently observed nebular forbidden emission lines is shown in
Figures~\ref{fig1}-\ref{fig4} for the four PNe.

\subsection{UV Line Measurements} \label{sec3.2}

\begin{figure}[t]
%\epsscale{0.6}
\plotone{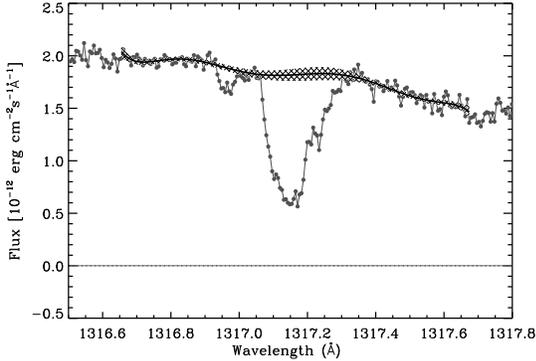}
\caption{A segment of the STIS spectrum that covers the \ion{Ni}{2}
line at 1317.217$\,$\AA\ for the central star in He2-138.  The large
feature centered at 1317.15$\,$\AA\ arises from foreground
interstellar material.  The small feature centered 1316.96$\,$\AA\ is
the one that is relevant to our study, since it arises from the
nebula.  The adopted continuum level is shown by a curved line with a
cross-hatched overlay that shows the 1$\sigma$ uncertainty in its
placement. \label{fig5} }%
\end{figure}

Table~\ref{tab3} lists the absorption lines that we measured in the
central star spectra of the four planetary nebulae.  Following the
name of each target in the subheaders of the table, we list the radial
velocity of the central star $v_*$ and the velocity interval that covers
the strongest absorption features that we identify as arising from the
planetary nebula shell.  Weaker features often spanned smaller
velocity intervals and the measurements of these lines, along with
those undetected, were taken over the more restricted ranges.

We defined the continuum levels by fitting Legendre polynomials to the
fluxes on either side of each line, using the methods refined by
\citet{SS92}.  In Figure~\ref{fig5} we present a portion of the UV
spectrum of the He2-138 central star which shows the final continuum
fit, together with the envelope defined by 1$\sigma$ excursions from
the fit, used in the determination of the absorption intensities in
the apparent optical depth analysis of the \ion{Ni}{2} \lam1317.217
line.

For each absorption feature, within the errors of the continuum
fitting there is an acceptable range for the reconstructed intensity
levels, and limits for this range defined the errors in the line
measurements attributable to continuum uncertainties.  For both the
equivalent width measurements and the evaluations of column densities
using the apparent optical depth (AOD) method (\S\ref{sec3.3}), we
combined these continuum uncertainties in quadrature with the
uncertainties due to photon-counting noise to arrive at a net error of
the quantity being measured.  In some instances, lines could be
measured twice because they appeared in two adjacent echelle orders.

While our principal objective was to obtain column densities for ions
in the nebular shells for comparison with emission line strengths from
the same ions, we nevertheless have included in Table~\ref{tab3}
measurements for absorption lines from ions for which there were no
emission lines detected in our ground-based spectra (\S\ref{sec4.2}).
We feel that it is prudent to include these lines for the benefit of
future more general studies of the relative abundances of atoms and
ions in the planetary nebulae shells.

Uncertain column densities result from either very weak lines that are
marginally detected or strongly saturated lines.  A few of the ions
have multiple lines with a range of f-values that provide for reliable
column densities.  We avoided lines that were so badly saturated that
their resulting lower limits for the column densities would be so much
lower than the actual values that they would have no real value for
any study.  For some species, e.g., \ion{Mg}{2}, only the strongest
line appeared above the noise; for such cases we could not measure the
weaker line. Using the same methods as for lines that were visible in
our spectra, we evaluated intensity upper limits within the wavelength
intervals where certain lines of interest might be expected, but which
were either marginally detectable or not visible at all.  Sometimes
these measurements yielded negative equivalent widths, although with
magnitudes comparable to or much less than the errors, and these
determinations are ultimately useful in providing upper limits for the
column densities (see footnote b to Table~\ref{tab3}).

\subsection{Absorption Column Densities} \label{sec3.3}

Column densities $N$ were derived by integrating over velocity the
apparent optical depths $\tau_a(v) = \ln [I_{cont}(v)/I(v)]$ and
evaluating the quantity
\begin{equation} \label{eq1}
N \equiv \int n_1 d\ell =  m_e c/(\pi e^2 f \lambda) \int \tau_a(v)
dv\,,
\end{equation}
where $n_1$ is the density of the ion in the lower level, and the
numerical value for the expression in front of the integral is
$3.77\times10^{14}$ cm$^{-2}$ (km s$^{-1}$)$^{-1}$ \citep{SS91,Je96}.
The results for all of the ions with reliable determinations are given
in Table~\ref{tab3}, and it should be emphasized that the column
densities refer only to those ions that occupy the lower level of the
transition.  Errors in the column densities may arise from three
different sources: (a) photon-counting noise, (b) errors in the
definition of the continuum level, and (c) errors in the adopted zero
intensity level.  If the random deviations of intensity arising from
statistical fluctuations in photon counts are expressed as
$\sigma_{I(v)}$, an approximation for the error in $\tau_a(v)$ is
simply
\begin{equation} \label{eq2}
\sigma_{\tau_a(v)} = \sigma_{I(v)} I_{cont}(v)/I(v)\,,
\end{equation}
which is reasonably accurate as long as the quantity is much less than
unity.  \citet{JT01} found that for a signal-to-noise ratio of about
20 at the continuum (which applies to nearly all of our spectral
lines) and a Gaussian error distribution, the approximation expressed
in equation~\ref{eq2} is good as long as $I(v)/I_{cont}(v)\ge0.15$.
We have indicated which lines appearing in Table~\ref{tab3} violate
this condition at the maximum level of absorption.  For these cases,
the upper error bounds may need to be increased to somewhat larger
values than those listed.

The errors in optical depth $\sigma_{\tau_a(v)}$ that arise from photon
counting are uncorrelated from one spectral element to the next, while
the systematic error arising from a misplacement of the continuum is
an effect that is usually coherent over the extent of an absorption
feature.  For this reason, the noise errors for successive spectral
elements were added together in quadrature before they were combined
as a group with the global uncertainty in line strength caused by
inaccuracies in the definition of the continuum level.  Since the
errors arising from photon counting and continuum misplacement are
uncorrelated, it is appropriate to add them together in quadrature.
At the bottoms of severely saturated lines, we found that the
intensities deviate from zero by less than 1\% of the continuum
intensity.  Thus, anomalies arising from errors in zero corrections
are insignificant compared to shortcomings of the approximation in
equation~\ref{eq2}.

Except for some strong lines of \ion{Si}{2} and \ion{Si}{2}* in the
spectrum of NGC~6543, the values of $N$ obtained for lines of
different strength generally agreed with each other.  In this
exception, the fact that the stronger lines yielded lower column
densities than the weaker ones for these two ground state levels
indicates that there are some unresolved saturated absorptions that
make all of the evaluations of $N$ using equation~\ref{eq1}
underestimate slightly the true value of $N$ for \ion{Si}{2}
\citep{Je96}.

\section{Emission-Line Analysis} \label{sec4}

\subsection{Optical Spectroscopy} \label{sec4.1}

\begin{deluxetable*}{
c    %lambda
c    %logflambda
c    %species
c    %w_lambda
c    %logN
}
\tablecolumns{5}
\tablewidth{380pt}
\tablecaption{Equivalent Widths and Column Densities\label{tab3}}
\tablehead{
\colhead{$\lambda$\tablenotemark{a}~~~} & \colhead{$\log
f\lambda$\tablenotemark{a}~~~} & \colhead{Species~~~~~~} &
\colhead{$W_\lambda$\tablenotemark{b}} & \colhead{$\log N$\tablenotemark{b}}\\
\colhead{(\AA)} & \colhead{} & \colhead{} & \colhead{(m\AA)} & \colhead{$({\rm
cm}^{-2}$)}\\
}
\startdata
\cutinhead{{\bf He2-138}~~($v_*=-47\,{\rm km~s}^{-1};~-73 < v_{\rm PN} < -40\,{\rm
km~s}^{-1}$)}
1656.928 &  2.392 &C I &$    28.6\pm  4.4 $&  12.97 ($+ 0.07, -0.08$)\\
1560.309 &  2.082 &C I &$     2.6\pm  1.8 $&  12.25 ($+ 0.22, -0.47$)\\
1277.245 &  2.037 &C I &$    11.2\pm  1.8 $&  13.00 ($+ 0.07, -0.08$)\\
1548.204 &  2.468 &C IV &$   32.5\pm  4.2 $&  12.99 ($+ 0.06, -0.07$)\\
1550.781 &  2.167 &C IV &$   15.7\pm  4.0 $&  12.93 ($+ 0.10, -0.13$)\\
1550.781 &  2.167 &C IV &$   11.8\pm  2.6 $&  12.79 ($+ 0.09, -0.11$)\\
1304.858 &  1.795 &O I* &$   75.1\pm  1.2 $&  14.49 ($+ 0.05,
-0.05$)\tablenotemark{c}\\
1306.029 &  1.795 &O I** &$  64.9\pm  1.4 $&  14.38 ($+ 0.04,
-0.04$)\tablenotemark{c}\\
1306.029 &  1.795 &O I** &$  67.7\pm  2.2 $&  14.20 ($+ 0.08,
-0.10$)\tablenotemark{c}\\
1239.925 & -0.106 &Mg II &$   5.0\pm  1.3 $&  14.79 ($+ 0.10, -0.14$)\\
1239.925 & -0.106 &Mg II &$   4.2\pm  1.1 $&  14.72 ($+ 0.10, -0.14$)\\
1670.789 &  3.463 &Al II &$  91.4\pm 20.5 $&  12.69 ($+ 0.12,
-0.17$)\tablenotemark{c,d}\\
1304.370 &  2.052 &Si II &$  88.9\pm  1.0 $&  14.57 ($+ 0.04,
-0.05$)\tablenotemark{c,d}\\
1309.276\tablenotemark{e} &  2.052 &Si II* &$ 69.6\pm  2.3 $&  13.98 ($+ 0.02,
-0.02$)\tablenotemark{f}\\
1309.276\tablenotemark{e} &  2.052 &Si II* &$ 71.5\pm  2.4 $&  13.98 ($+ 0.02,
-0.02$)\tablenotemark{f}\\
1152.818 &  2.451 &P II &$   31.2\pm  3.8 $&  13.28 ($+ 0.09,
-0.11$)\tablenotemark{f}\\
1301.874 &  1.219 &P II &$    4.1\pm  1.1 $&  13.36 ($+ 0.11, -0.14$)\\
1153.995 &  2.331 &P II** & $11.2\pm  4.5 $&  12.83 ($+ 0.14, -0.22$)\\
1295.653 &  2.052 &S I &$     1.6\pm  1.6 $&  $<12.58$\\
1250.578 &  0.832 &S II &$   74.9\pm  2.0 $&  15.49 ($+ 0.08,
-0.09$)\tablenotemark{c}\\
1250.578 &  0.832 &S II &$   71.7\pm  1.2 $&  15.42 ($+ 0.10,
-0.14$)\tablenotemark{c}\\
1190.203 &  1.449 &S III &$  73.6\pm  2.3 $&  14.76 ($+ 0.12,
-0.16$)\tablenotemark{f}\\
1190.203 &  1.449 &S III &$  70.0\pm  1.6 $&  14.70 ($+ 0.02,
-0.02$)\tablenotemark{f}\\
1201.729 &  0.626 &S III**&$  16.3\pm  3.2$&  14.63 ($+ 0.08, -0.10$)\\
1197.184 &  2.414 &Mn II &$   9.4\pm  2.5 $&  12.66 ($+ 0.11, -0.15$)\\
1197.184 &  2.414 &Mn II &$  10.8\pm  2.2 $&  12.68 ($+ 0.08, -0.10$)\\
1199.391 &  2.308 &Mn II &$   6.2\pm  2.5 $&  12.53 ($+ 0.14, -0.22$)\\
1608.451 &  1.968 &Fe II &$  55.0\pm  2.3 $&  13.95 ($+ 0.03,
-0.04$)\tablenotemark{f}\\
1317.217 &  1.876 &Ni II &$   4.1\pm  1.3 $&  12.70 ($+ 0.12, -0.16$)\\
1317.217 &  1.876 &Ni II &$   5.3\pm  1.1 $&  12.81 ($+ 0.09, -0.11$)\\
1237.059 &  3.183 &Ge II &$   9.7\pm  1.3 $&  11.86 ($+ 0.06, -0.06$)\\
1237.059 &  3.183 &Ge II &$  11.6\pm  1.2 $&  11.94 ($+ 0.05, -0.05$)\\
1235.838 &  2.402 &Kr I &$    3.4\pm  1.9 $&  12.14 ($+ 0.18, -0.33$)\\
1235.838 &  2.402 &Kr I &$    4.6\pm  1.2 $&  12.25 ($+ 0.10, -0.14$)\\
\cutinhead{{\bf NGC~246}~~($v_*=-46\,{\rm km~s}^{-1};~-95 < v_{\rm PN} < -60\,{\rm
km~s}^{-1}$)}
1277.245 &  2.037 &C I &$     0.2\pm  1.0 $& $<12.29$\\
1334.532 &  2.234 &C II &$   12.9\pm  1.1 $&  12.84 ($+ 0.04, -0.04$)\\
1334.532 &  2.234 &C II &$   14.3\pm  1.0 $&  12.89 ($+ 0.03, -0.03$)\\
1335.708 &  2.234 &C II* &$  15.8\pm  1.0 $&  12.93 ($+ 0.03, -0.03$)\\
1199.550 &  2.199 &N I &$    -0.7\pm  1.0 $& $<12.04$\\
1302.169 &  1.796 &O I &$   4.3\pm  2.1 $&  12.80 ($+ 0.17,
-0.28$)\tablenotemark{g}\\
1302.169 &  1.796 &O I &$   0.6\pm  1.7 $& $<12.78$\tablenotemark{g}\\
1304.858 &  1.795 &O I* &$   -1.7\pm  1.3 $& $<12.44$\\
1260.422 &  3.171 &Si II &$   6.5\pm  2.6 $&  11.61 ($+ 0.14, -0.22$)\\
1264.738 &  3.125 &SiII* &$  -0.3\pm  1.0 $& $<11.14$\\
1206.500 &  3.293 &Si III &$ 49.7\pm  3.5 $&  12.52 ($+ 0.03, -0.03$)\\
1393.760 &  2.854 &Si IV &$ 145.1\pm  2.3 $&  13.50 ($+ 0.01,
-0.01$)\tablenotemark{f}\\
1402.773 &  2.552 &Si IV &$  97.6\pm  1.3 $&  13.51 ($+ 0.01, -0.01$)\\
1259.518 &  1.320 &S II &$    3.5\pm  2.2 $&  13.19 ($+ 0.20, -0.40$)\\
1190.203 &  1.449 &S III &$   3.4\pm  1.3 $&  13.08 ($+ 0.14, -0.21$)\\
1194.058 &  1.325 &S III* &$ -1.7\pm  2.5 $& $<13.32$\\
1317.217 &  1.876 &Ni II &$   2.1\pm  1.8 $&  12.40 ($+ 0.26, -0.71$)\\
1370.132 &  1.906 &Ni II &$  -0.6\pm  1.9 $& $<12.61$\\
\cutinhead{{\bf NGC~6543}~~($v_*=-66\,{\rm km~s}^{-1};~-103 < v_{\rm PN} <
-74\,{\rm km~s}^{-1}$)}
1277.245  &  2.037 &C I &$   1.1\pm  1.4 $& $<12.52$\\
1200.223  &  2.018 &N I &$   2.0\pm  1.7 $&  12.29 ($+ 0.25, -0.69$)\\
1304.858  &  1.795 &O I* &$   1.7\pm  1.3 $&  12.39 ($+ 0.25, -0.65$)\\
1306.029  &  1.795 &O I** &$  -0.3\pm  1.8 $& $<12.73$\\
1306.029  &  1.795 &O I** &$  -1.4\pm  1.3 $& $<12.47$\\
1670.789  &  3.463 &Al II &$   6.5\pm  4.5 $& 11.22 ($+0.22,-0.48$)\\  
1260.422  &  3.171 &Si II &$  85.9\pm  1.6 $&  13.11 ($+ 0.02,
-0.02$)\tablenotemark{c}\\
1526.707  &  2.307 &Si II &$  35.7\pm  2.0 $& 13.23 ($+0.03,-0.03$)\\ 
1304.370  &  2.052 &Si II &$  19.8\pm  1.4 $&  13.25 ($+ 0.03, -0.03$)\\
1264.738  &  3.125 &Si II* &$  88.8\pm  0.9 $&  13.17 ($+ 0.02,
-0.02$)\tablenotemark{c}\\
\enddata
\end{deluxetable*}

\addtocounter{table}{-1}
\begin{deluxetable*}{
c    %lambda
c    %logflambda
c    %species
c    %w_lambda
c    %logN
}
\tablecolumns{5}
\tablewidth{380pt}
\tablecaption{(continued)}
\tablehead{
\colhead{$\lambda$\tablenotemark{a}~~~} & \colhead{$\log
f\lambda$\tablenotemark{a}~~~} & \colhead{Species~~~~~~} &
\colhead{$W_\lambda$\tablenotemark{b}} & \colhead{$\log N$\tablenotemark{b}}\\
\colhead{(\AA)} & \colhead{} & \colhead{} & \colhead{(m\AA)} & \colhead{$({\rm
cm}^{-2}$)}\\
}
\startdata
\cutinhead{{\bf NGC~6543}~~($v_*=-66\,{\rm km~s}^{-1};~-103 < v_{\rm PN} <
-74\,{\rm km~s}^{-1}$)}
1533.432  &  2.307 &Si II* &$  43.5\pm  2.9 $& 13.33 ($+0.03,-0.03$)\\ 
1265.002  &  2.171 &Si II* &$  33.2\pm  1.5 $&  13.39 ($+ 0.02, -0.02$)\\
1309.276  &  2.052 &Si II* &$  22.6\pm  1.0 $&  13.31 ($+ 0.02, -0.02$)\\
1152.818  &  2.451 &P II &$   5.8\pm  6.2 $& $<12.81$\\
1259.518  &  1.320 &S II &$   9.6\pm  2.4 $&  13.65 ($+ 0.10, -0.12$)\\
1253.805  &  1.136 &S II &$   6.3\pm  1.3 $&  13.64 ($+ 0.08, -0.10$)\\
1190.203  &  1.449 &S III &$  93.2\pm  1.4 $&  14.92 ($+ 0.29,
-1.33$)\tablenotemark{c}\\
1194.058  &  1.325 &S III* &$  87.2\pm  1.7 $&  14.96 ($+ 0.02,
-0.03$)\tablenotemark{f}\\
1194.058  &  1.325 &S III* &$  86.5\pm  1.6 $&  14.95 ($+ 0.02,
-0.02$)\tablenotemark{f}\\
1197.184  &  2.414 &Mn II &$   1.3\pm  2.1 $& $<12.32$\\
1199.391  &  2.308 &Mn II &$   0.2\pm  2.2 $& $<12.38$\\
1608.451  &  1.968 &Fe II &$   0.1\pm  4.0 $& $<12.84$\\   
1317.217  &  1.876 &Ni II &$  -0.7\pm  1.6 $& $<12.56$\\
1370.132  &  1.906 &Ni II &$  -1.2\pm  2.2 $& $<12.64$\\
%\tablebreak
\cutinhead{{\bf Tc 1}~~($v_*=-83\,{\rm km~s}^{-1};~-128 < v_{\rm PN} < -86\,{\rm
km~s}^{-1}$)}
1560.309  &  2.082 &C I &$  -3.2\pm  4.1 $& $<12.65$\\
1277.245  &  2.037 &C I &$  -1.9\pm  2.4 $& $<12.54$\\
1199.550  &  2.199 &N I &$  -2.1\pm  3.1 $& $<12.53$\\
1302.169  &  1.796 &O I &$  27.1\pm  2.4 $&  13.63 ($+ 0.04,
-0.04$)\tablenotemark{h}\\
1306.029  &  1.795 &O I** &$   0.5\pm  2.0 $& $<12.83$\\
1239.925  & -0.106 &Mg II &$   0.2\pm  3.2 $& $<14.94$\\
1239.925  & -0.106 &Mg II &$   7.2\pm  3.8 $&  14.94 ($+ 0.18, -0.32$)\\
1670.787  &  3.463 &Al II &$  74.8\pm 22.9 $&  12.34 ($+ 0.12, -0.16$)\\
1260.422  &  3.171 &Si II &$ 110.7\pm  2.1 $&  13.22 ($+ 0.02,
-0.02$)\tablenotemark{c,d}\\
1193.290  &  2.842 &Si II &$  66.9\pm  3.7 $&  13.17 ($+ 0.03, -0.03$)\\
1190.416  &  2.541 &Si II &$  38.1\pm  4.4 $&  13.15 ($+ 0.05, -0.05$)\\
1304.370  &  2.052 &Si II &$  22.3\pm  1.2 $&  13.29 ($+ 0.02, -0.03$)\\
1264.738  &  3.125 &Si II* &$  99.8\pm  1.7 $&  13.18 ($+ 0.02,
-0.02$)\tablenotemark{c}\\
1265.002  &  2.171 &Si II* &$  22.9\pm  2.4 $&  13.20 ($+ 0.04, -0.05$)\\
1309.276  &  2.052 &Si II* &$  22.8\pm  1.7 $&  13.29 ($+ 0.03, -0.03$)\\
1152.818  &  2.451 &P II &$   7.3\pm  6.9 $&  12.52 ($+ 0.25, -0.66$)\\
1259.518  &  1.320 &S II &$  10.3\pm  2.0 $&  13.67 ($+ 0.08, -0.09$)\\
1190.203  &  1.449 &S III &$  79.9\pm  2.6 $&  14.65 ($+ 0.02, -0.02$)\\
1194.058  &  1.325 &S III* &$  49.5\pm  5.6 $&  14.46 ($+ 0.05, -0.06$)\\
1194.058  &  1.325 &S III* &$  48.4\pm  6.2 $&  14.46 ($+ 0.06, -0.07$)\\
1197.184  &  2.414 &Mn II &$  25.8\pm  4.8 $&  13.06 ($+ 0.08, -0.09$)\\
1197.184  &  2.414 &Mn II &$  25.9\pm  2.9 $&  13.05 ($+ 0.05, -0.05$)\\
1608.451  &  1.968 &Fe II &$  -2.9\pm  8.5 $& $<13.12$\\
1317.217  &  1.876 &Ni II &$   0.4\pm  1.4 $& $<12.60$\\
1237.059  &  3.183 &Ge II &$  -5.3\pm  3.2 $& $<11.43$\\
1237.059  &  3.183 &Ge II &$  -2.4\pm  3.8 $& $<11.63$\\
1235.838  &  2.402 &Kr I &$   9.0\pm  2.6 $&  12.53 ($+ 0.11, -0.15$)\\
\enddata

\tablenotetext{a}{Wavelengths and line strengths from
\citet{Mo00,Mo03}, except for the $f$-values of \ion{Ni}{2}, for which
we have adopted the values measured by \citet{JT01}.  Transitions for
individual species are arranged according to decreasing line strength.
This was done in order to make it easy to identify trends (strong
lines indicating smaller $N$ than weak ones) that signify possible
unresolved saturated components that could lead to underestimates of
column density using the AOD method \citep{SS91,Je96}.  Duplicate
entries signify independent measurements made in different echelle
orders.}
%
% Note "b" below is referenced in the text.  If it's changed, be sure
% to also change the text accordingly.
%
\tablenotetext{b}{{Listed} errors represent $\pm 1\sigma$ deviations and
include uncertainties caused by both photoevent statistical fluctuations
and continuum uncertainties, combined in quadrature.  When a measurement
of $W_\lambda$ yields a value that is below the calculated $1\sigma$
error in $W_\lambda$, we state the formal measurement of $W_\lambda$ and
its error, but then we follow with an evaluation of a $2\sigma$ upper
confidence bound for the real $W_\lambda$ using the method of
\citet{Ma92} for interpreting marginal detections (or nondetections) of
quantities that are known not to ever be negative.  The listed upper
limit for $N$ is calculated from this $W_\lambda$ limit using the
formula for weak lines (i.e., the linear part of the curve of
growth).}
\tablenotetext{c}{The line is strongly saturated (central optical depth
$\tau_0 \gtrsim 4$), but without a flat bottom that would signify severe
saturation.  The formal errors listed here may not accurately reflect
the true errors.  In our apparent optical depth integrations, occasional
deviations in $\tau_a(v)$ that exceeded 5.0 were simply set to equal to
5.0.}
\tablenotetext{d}{The right-hand portion of the profile is partly
blended with the left-hand portion of the absorption arising from
foreground material in the general interstellar medium.  Thus, the
errors could be somewhat larger than those derived formally (and stated
here).}
\tablenotetext{e}{\ion{Si}{2}* was also recorded at 1533.4$\,$\AA, but
this line is strong enough to have a small portion of its profile
strongly saturated.  Since our recording of the 1309.3$\,$\AA\ feature
is of excellent quality (and appeared in two orders), we decided not
to measure the stronger line.}
\end{deluxetable*}

\addtocounter{table}{-1}
\begin{deluxetable*}{
c    %lambda
c    %logflambda
c    %species
c    %w_lambda
c    %logN
}
\tablecolumns{5}
\tablewidth{380pt}
\tablecaption{(continued)}
%\tablehead{
%\colhead{$\lambda$\tablenotemark{a}~~~} & \colhead{$\log
%f\lambda$\tablenotemark{a}~~~} & \colhead{Species~~~~~~} &
%\colhead{$W_\lambda$\tablenotemark{b}} & \colhead{$\log N$\tablenotemark{b}}\\
%\colhead{(\AA)} & \colhead{} & \colhead{} & \colhead{(m\AA)} & \colhead{$({\rm
%cm}^{-2}$)}\\
%}
\startdata
\enddata 
\tablenotetext{f}{At the bottom of the line, the intensity relative to
the continuum is less than 0.15.  For our representative $S/N=20$ (at
the continuum), the approximation given in equation~\ref{eq2} starts
to become inaccurate.  For this reason, the upper bound for $\log N$
should be increased slightly beyond the value listed here.}
\tablenotetext{g}{Absorption by the 1301.874$\,$\AA\ transition of
\ion{P}{2} caused by foreground gas interferes with the \ion{O}{1}
feature.  However, we could compensate for this by dividing the
spectrum by the \ion{P}{2} profile at 1152.818$\,$\AA\ after its
strength had been reduced to reflect the fact that the
1301.874$\,$\AA\ line is weaker.  (All intensities in the strong
profile were taken to a power equal to the ratio of the lines' values
of $f\lambda$.)}
\tablenotetext{h}{The right-hand portion of this profile is partly
blended with the left-hand side of an absorption arising from the
1301.874$\,$\AA\ transition of \ion{P}{2} created by the foreground
gas.}
\end{deluxetable*}

\begin{figure*}[t]
\epsscale{0.65}
\plottwo{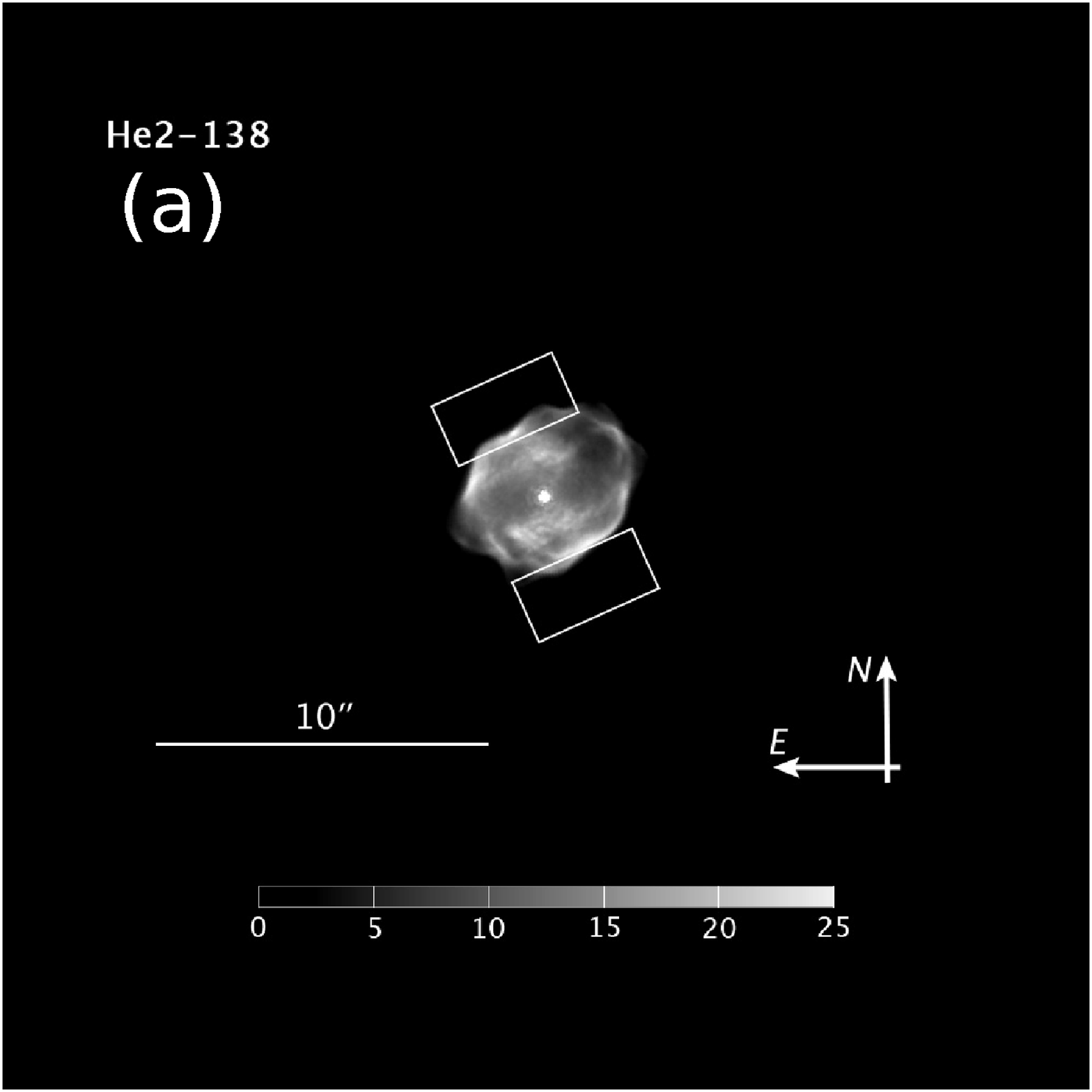}{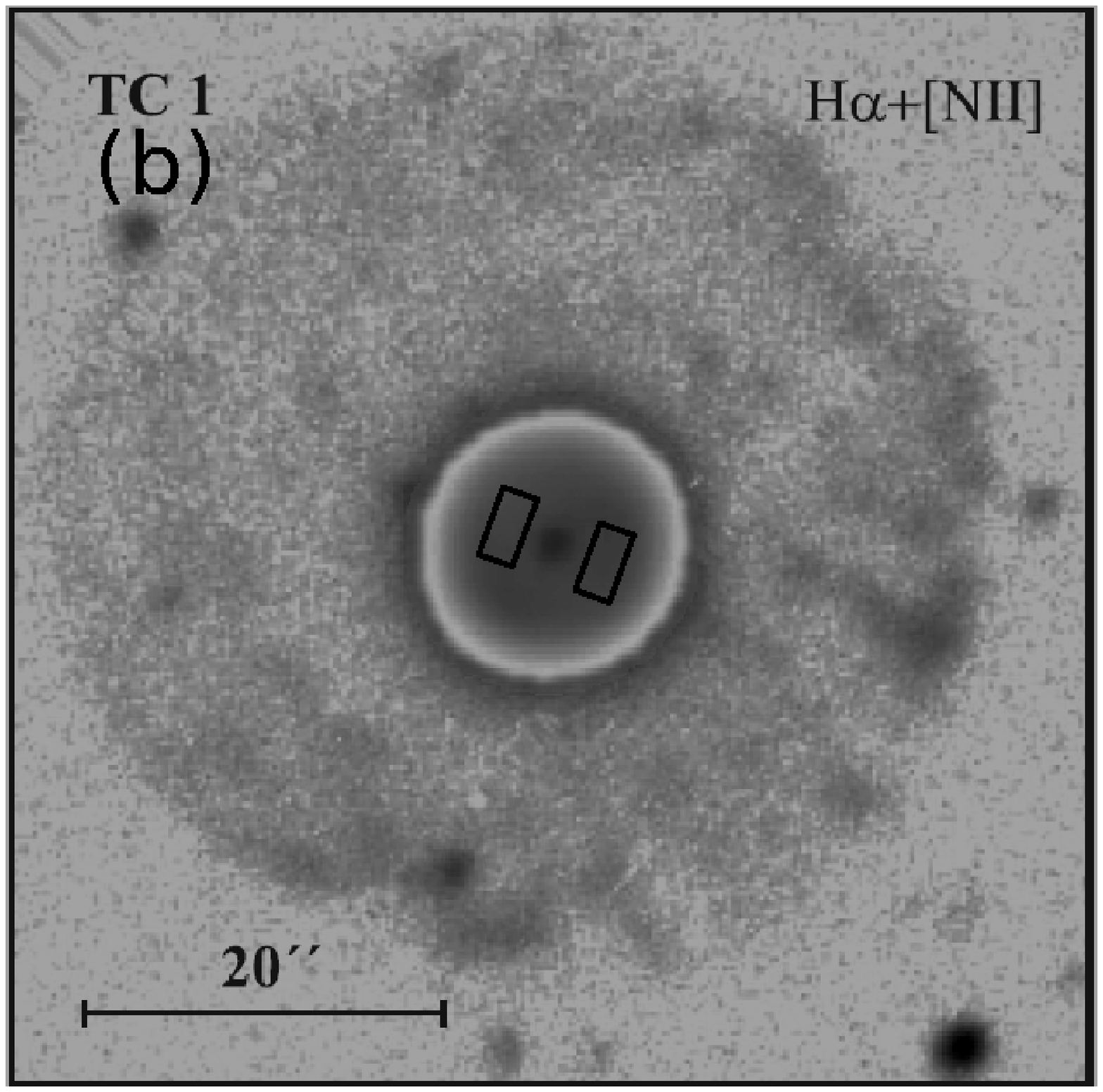}
\epsscale{0.32}
\plotone{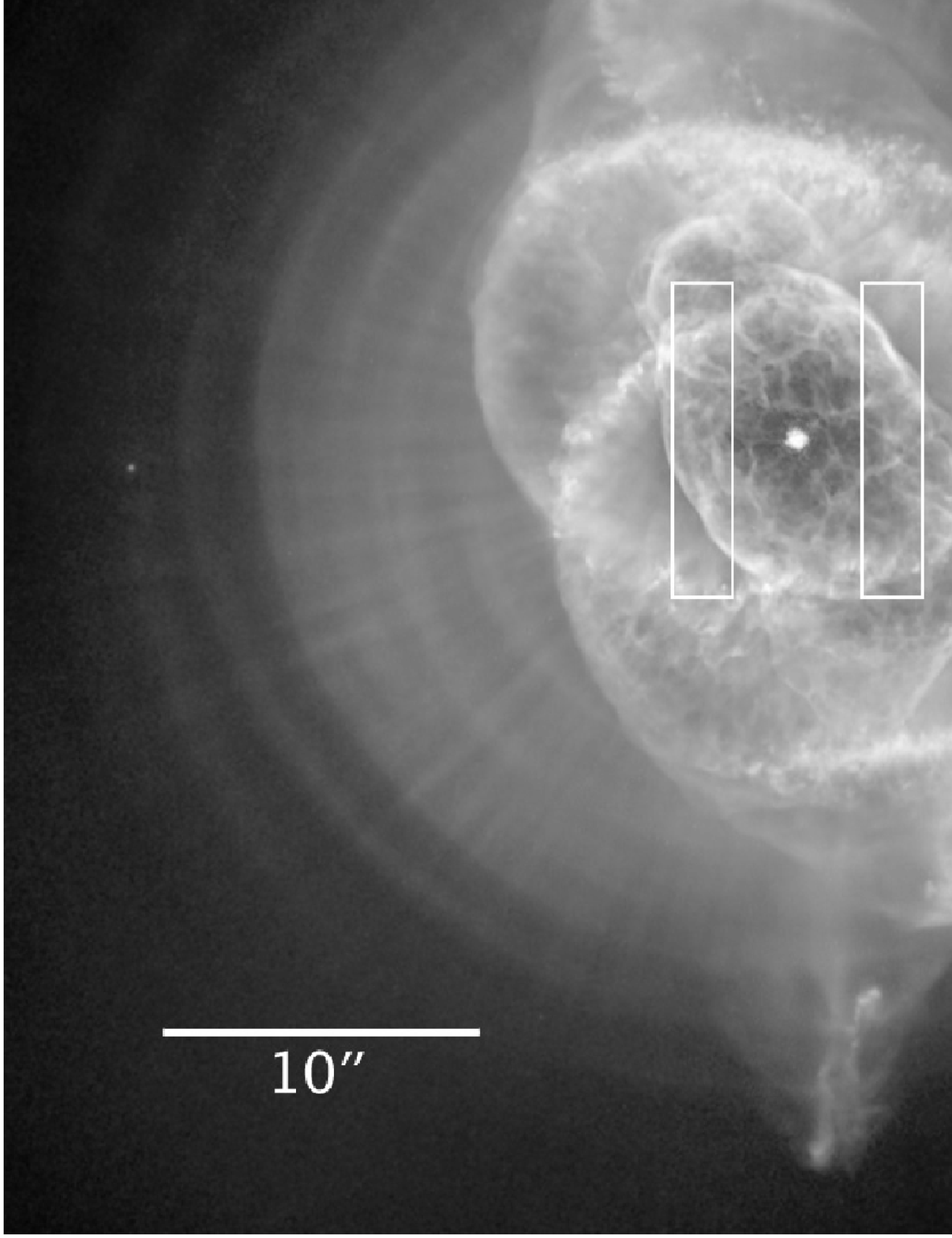}
\caption{Images of program PNe showing locations, orientations, and
size of slits for (a) He2-138, (b) Tc1, and (c) NGC~6543.  In all
figures North is up and East is to left. Panels (a) and (c) are HST
images.  The shell emission sampled by our spectra would be
represented by these images convolved with a 1--1.3 arcsec
PSF. \label{fig6}}
\end{figure*}

The four PNe studied here were observed in the visible with
ground-based telescopes to measure intensities of the nebular emission
lines.  The lower surface brightnesses of the nebular shells compared
with the central stars dictated that the nebular spectra sight lines
be positioned no closer than 2 arcsec from the central star in order
to avoid unacceptable levels of contamination by scattered light from
the brighter star.  However, our goal was to estimate the intensity
directly along the line of sight to the central star since this is the
position of the gas that produces the absorption lines.  Our approach
was to take spectra in two positions symmetrically placed on either
side of the central star and to then average together the two spectra
after they had been flux calibrated.  The flux average serves to
compensate for surface brightness fluctuations, but does not represent
the flux along the sight line to the central star if there is a radial
gradient in surface brightness away from the central star due to the
three-dimensional structure of the shell.

Initially, a reconnaissance was carried out in 2004 at low spectral
resolution (300--500 km s$^{-1}$ FWHM) using the Gold Spectrograph on
the 2.1m telescope at Kitt Peak National Observatory (KPNO) to observe
NGC~246 and NGC~6543, and the Wide Field CCD (WFCCD) camera on the du
Pont 2.5m telescope at Las Campanas Observatory (LCO) to observe
He2-138, NGC~246, and Tc 1.  We found that the emission lines in NGC
246 are much too faint for accurate spectrophotometric measurements of
any of the weaker emission lines; only a few of the very strongest
lines such as [\ion{O}{3}] $\lambda$5007 were visible even in long
exposures. This object was therefore removed from our program, but we
provide the UV information from the central star spectrum in this
paper because of its potential use for abundance studies.  The visible
spectra of the nebulae showed that line blending dictated the need for
much higher resolution data to show many of the weaker emission lines
of interest in the other three objects. This led us to obtain 15--20
km s$^{-1}$ FWHM resolution echelle spectra in 2005 with the echelle
spectrographs on the 4m Mayall Telescope at KPNO and the 2.5m du Pont
Telescope at LCO.

\subsubsection{Observations at LCO} \label{sec4.1.1}

The LCO echelle spectrograph uses a prism as a cross disperser and
covers the wavelength range $\lambda\lambda$3480--10,150\AA\ in 64
orders, with increasingly large wavelength gaps between orders beyond
$\lambda$8000\AA.  On each of the two nights 8, 9 June 2005 UT, we
observed He2-138 and Tc 1 in two different slit positions
symmetrically placed on either side of the central star, using a 2
arcsec wide $\times$ 4 arcsec long slit and offsetting at an angle so
that the slit would include the brightest part of the nebulae. The
slit orientations were at right angles to the directions of the
offsets.  The two slit positions for Tc 1 were 1.0 arcsec S, 2.7
arcsec W of the central star, and 1.0 arcsec N, 2.7 arcsec E.  For
He2-138, the two slit positions were 2.7 arcsec N, 1.25 arcsec E and
2.7 arcsec S, 1.25 arcsec W of the central star.  The slit positions
used for our observations are shown in Figure~\ref{fig6}, overlayed on
the best available images that we could find for these objects.  

For Tc 1 the spectra were taken well inside the outer edge of the
nebula.  However, for the smaller He2-138 we were not able to
simultaneously avoid the scattering effects of the central star and
sample the shell well inside its outer edge, so the slit had to be
placed near the outer edge of the nebula.  Figure 6a shows a Hubble
Telescope image of He2-138 with the spectrograph slit positions
superposed.  Atmospheric seeing resulted in shell emission filling the
central portion of the slit.  We detected flux along the central 3
arcsec of the slit, but variations in tracking and seeing caused the
intensity to vary among the exposures.  These variations make our
determinations of the emission measure for this object considerably
less certain than for the other two PNe.  The spectral resolution was
15 km s$^{-1}$ FWHM over most of the range, but degraded to 20 km
s$^{-1}$ at the extreme ends.  We extracted spectra from each
position, and after applying the proper flux calibration averaged the
two extracted spectra together.  We added together seven 1200 sec
exposures at each slit position over two nights to measure the weak
lines, and used pairs of 30 or 60 sec exposures at each position to
measure the strong emission lines which would otherwise be saturated.

\begin{figure}[t]
\epsscale{1}
\plotone{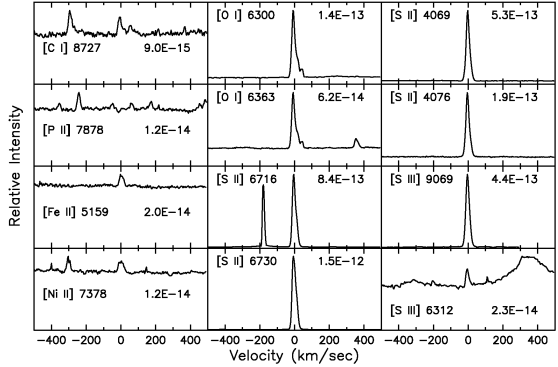}
\caption{Nebular emission line profiles from the LCO echelle spectrum
of He2-138.  The panels show the forbidden emission lines listed in
Table~\ref{tab7} that were used to determine forbidden line
abundances.  The corresponding lines are shown at zero velocity in
each panel.  The linear vertical scale is F$_\lambda$ in units of erg
cm$^{-2}$ s$^{-1}$ \AA$^{-1}$ with the bottom abscissa of each panel
corresponding to zero flux and the top corresponding to the value of
F$_\lambda$ printed inside the panel. \label{fig7}}
\end{figure}  

\begin{figure}
\plotone{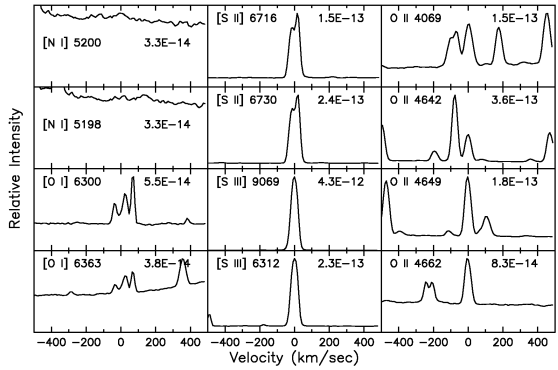}
\caption{Emission line profiles from the KPNO echelle spectrum of
NGC~6543. The panels and scaling are as in Figure~\ref{fig7}, except
that the right-hand column shows four of the stronger \ion{O}{2}
recombination lines from Table~\ref{tab10}.  The stronger red
[\ion{O}{1}] emission components at +70 km s$^{-1}$ are atmospheric
airglow lines. \label{fig8}
}
\end{figure}

\begin{figure}[th]
\plotone{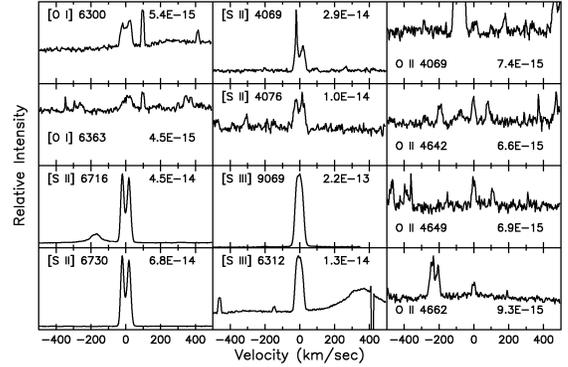}
\caption{Emission line profiles from the LCO echelle spectrum of Tc
  1. The panels and scaling are as in Figure~\ref{fig7}, with the
  narrow [\ion{O}{1}] components at +100 km s$^{-1}$ again due to
  atmospheric [\ion{O}{1}] emisison.  \label{fig9} }
\end{figure}

Only the second night was photometric so we used observations of the
standard stars HR 4468, HR 4963 and HR 5501 \citep{Ha94} made through
a 8$\times$8 arcsec slit on the second night to calibrate both nights.
We applied this calibration in three steps.  We first added together
the raw counts from all of the different long exposures, which is the
optimal weighting for detecting weak lines, and then flux calibrated
that spectrum using a mean airmass.  This gave a high signal-to-noise
ratio spectrum that was affected by a wavelength-independent
attenuation and by a slight error in the wavelength dependence of the
extinction correction.  The second step was to properly flux calibrate
a single long exposure in each slit position on the photometric night
using the correct airmass.  Finally, we measured the fluxes of the
same intermediate strength emission lines in both spectra, and used
the flux ratios to correct the high S/N spectrum to match the flux
scale of the single well-calibrated spectrum.  This procedure provides
the best calibration for our observing circumstances and results in
absolute spectrophotometry accurate to better than 8 percent over the
whole wavelength range.

\subsubsection{Observations at KPNO} \label{sec4.1.2}

\begin{deluxetable*}{clcc@{\extracolsep{0.2in}}c@{\extracolsep{0.15in}}c@{\extracolsep{0.2in}}c@{\extracolsep{0.15in}}c}
\tablecaption{Optical Forbidden Emission Line Fluxes \label{tab4}}
\tablewidth{6.75in}
%\tabletypesize{\footnotesize}
%\rotate
\tablehead{
 & &
\multicolumn{2}{c}{He2-138} &
\multicolumn{2}{c}{NGC~6543} &
\multicolumn{2}{c}{Tc~1} \\ \cline{3-4}\cline{5-6}\cline{7-8}
\multicolumn{1}{c}{Species} &
\multicolumn{1}{c}{\lam (\AA)} &
\colhead{$F$\tablenotemark{a,b}} & %(erg cm$^{-2}$ s$^{-1}$)} & 
\colhead{$F_c$\tablenotemark{c}} & %(erg cm$^{-2}$ s$^{-1}$)} &
\colhead{$F$} & %(erg cm$^{-2}$ s$^{-1}$)} & 
\colhead{$F_c$} & %(erg cm$^{-2}$ s$^{-1}$)} &
\colhead{$F$} & %(erg cm$^{-2}$ s$^{-1}$)} & 
\colhead{$F_c$} %(erg cm$^{-2}$ s$^{-1}$)}
}
\startdata
\ion{H}{1}  & 4861 & 3.49(-12)   & 1.26(-11)   & 1.51(-11) & 2.10(-11) & 1.58(-12)   & 3.37(-12)   \\
\ion{C}{1}  & 4622 & 2.76(-16):  & 1.07(-15):  & \nodata   & \nodata   & \nodata     & \nodata     \\
\ion{C}{1}  & 8727 & 1.63(-15)   & 2.86(-15)   & \nodata   & \nodata   &$<$1.97(-16) &$<$2.76(-16) \\
\ion{C}{1}  & 9824 & 1.16(-15):  & 1.84(-15):  & \nodata   & \nodata   & \nodata     & \nodata     \\
\ion{C}{1}  & 9850 & \nodata     & \nodata     & \nodata   & \nodata   &$<$9.74(-16) &$<$1.28(-15) \\
\ion{N}{1}  & 5198 & 3.08(-14)   & 9.97(-14)   & 4.37(-15):& 5.92(-15):& 3.79(-16):  & 7.61(-16):  \\
\ion{N}{1}  & 5200 & 1.88(-14)   & 6.09(-14)   & 3.92(-15):& 5.30(-15):& 4.04(-16):  & 8.10(-16):  \\
\ion{N}{2}  & 5755 & 3.60(-14)   & 1.01(-13)   & 4.70(-14) & 6.13(-14) & 2.00(-14)   & 3.68(-14)   \\
\ion{N}{2}  & 6548 & 3.08(-12)   & 7.37(-12)   & 7.66(-13) & 9.59(-13) & 6.17(-13)   & 1.04(-12)   \\
\ion{N}{2}  & 6583 & 9.38(-12)   & 2.23(-11)   & 2.55(-12) & 3.18(-12) & 1.92(-12)   & 3.21(-12)   \\
\ion{O}{1}  & 5577 & 6.33(-16)   & 1.84(-15)   & \nodata   & \nodata   & 6.61(-17):  & 1.25(-16):  \\
\ion{O}{1}  & 6300 & 7.36(-14)   & 1.85(-13)   & 2.12(-14) & 2.69(-14) & 2.45(-15)   & 4.24(-15)   \\
\ion{O}{1}  & 6364 & 2.66(-14)   & 6.60(-14)   & 7.03(-15) & 8.82(-15) & 9.68(-16)   & 1.66(-15)   \\
\ion{O}{2}  & 3726 & 4.51(-13)   & 2.25(-12)   & 1.40(-12) & 2.12(-12) & 1.68(-12)   & 4.36(-12)   \\
\ion{O}{2}  & 3729 & 2.00(-13)   & 9.97(-13)   & 6.41(-13) & 9.70(-13) & 1.11(-12)   & 2.90(-12)   \\
\ion{O}{2}  & 7320 & 5.02(-12)   & 1.06(-13)   & 1.74(-13) & 2.10(-13) & 1.21(-13)   & 1.89(-13)   \\
\ion{O}{2}  & 7330 & 4.74(-14)   & 9.97(-14)   & 1.79(-13) & 2.17(-13) & 1.02(-13)   & 1.59(-13)   \\
\ion{O}{3}  & 4363 &$<$6.29(-16) &$<$2.64(-15) & 3.73(-13) & 5.40(-13) & 7.97(-15)   & 1.87(-14)   \\
\ion{O}{3}  & 4959 & 5.72(-16)   & 1.99(-15)   & 4.15(-11) & 5.73(-11) & 6.72(-13)   & 1.41(-12)   \\ 
\ion{O}{3}  & 5007 & 2.72(-15)   & 9.34(-15)   & 1.24(-10) & 1.71(-10) & 2.00(-12)   & 4.17(-12)   \\
\ion{P}{2}  & 4669 &$<$6.29(-16) &$<$2.40(-15) & \nodata   & \nodata   &$<$2.07(-16) &$<$4.58(-16) \\
\ion{P}{2}  & 7876 & 1.20(-16)   & 2.34(-16)   & \nodata   & \nodata   &$<$5.98(-17) &$<$8.88(-17) \\
\ion{S}{1}  & 4589 &$<$8.88(-16) &$<$3.48(-15) & \nodata   & \nodata   &$<$2.07(-16) &$<$4.66(-16) \\
\ion{S}{1}  & 7725 & \nodata     & \nodata     & \nodata   & \nodata   &$<$7.99(-17) &$<$1.20(-16) \\
\ion{S}{2}  & 4069 & 1.23(-13)   & 5.64(-13)   & 6.84(-14) & 1.01(-13) & 8.36(-15)   & 2.06(-14)   \\
\ion{S}{2}  & 4076 & 4.47(-14)   & 2.04(-13)   & 1.24(-14) & 1.83(-14) & 2.40(-15)   & 5.91(-15)   \\
\ion{S}{2}  & 6716 & 4.69(-13)   & 1.09(-12)   & 1.38(-13) & 1.71(-13) & 4.50(-14)   & 7.43(-14)   \\
\ion{S}{2}  & 6731 & 9.77(-13)   & 2.26(-12)   & 2.54(-13) & 3.15(-13) & 7.15(-14)   & 1.18(-13)   \\
\ion{S}{3}  & 6312 & 2.24(-15)   & 5.60(-15)   & 1.71(-13) & 2.17(-13) & 9.22(-15)   & 1.59(-14)   \\
\ion{S}{3}  & 9069 & 3.44(-13)   & 5.85(-13)   & 4.91(-12) & 5.63(-12) & 3.19(-13)   & 4.38(-13)   \\
\ion{Cl}{3} & 5518 & 4.57(-16)   & 1.35(-15)   & 6.54(-14) & 8.65(-14) & 5.06(-15)   & 9.62(-15)   \\
\ion{Cl}{3} & 5538 & 7.90(-16)   & 2.32(-15)   & 9.27(-14) & 1.22(-13) & 5.39(-15)   & 1.02(-14)   \\
\ion{Ar}{3} & 5192 & \nodata     & \nodata     & 1.05(-14) & 1.42(-14) & 5.29(-16)   & 1.06(-15)   \\ 
\ion{Ar}{3} & 7136 & 2.78(-15)   & 6.02(-15)   & 3.56(-12) & 4.35(-12) & 1.53(-13)   & 2.42(-13)   \\
\ion{Ar}{3} & 7751 & 3.75(-16)   & 7.34(-16)   & 8.30(-13) & 9.89(-13) & 3.85(-14)   & 5.78(-14)   \\
\ion{Ar}{4} & 4711 &  \nodata    & \nodata     & 1.77(-13) & 2.49(-13) & \nodata     &  \nodata    \\
\ion{Ar}{4} & 4740 &  \nodata    &  \nodata    & 2.08(-13) & 2.92(-13) & \nodata     &  \nodata    \\
\ion{Fe}{2} & 4244 &$<$1.44(-15) &$<$6.28(-15) & 1.10(-15):& 1.61(-15):&$<$3.12(-16) &$<7$.47(-16) \\
\ion{Fe}{2} & 4359 & 3.56(-15)   & 1.50(-14):  & 1.39(-15):& 2.01(-15):&$<$2.08(-16) &$<$4.87(-16) \\
\ion{Fe}{2} & 5159 & 1.58(-15)   & 5.17(-15):  & \nodata   & \nodata   & \nodata     & \nodata     \\
\ion{Fe}{2} & 8617 & 2.75(-15)   & 4.91(-15):  & \nodata   & \nodata   &$<$1.98(-16) &$<$2.78(-16) \\
\ion{Ni}{2} & 6667 & 1.49(-15):  & 3.48(-15):  & 4.69(-16):& 5.83(-16):& 2.29(-16):  & 3.80(-16):  \\
\ion{Ni}{2} & 7376 & 1.71(-15):  & 3.58(-15):  & \nodata   & \nodata   &$<$1.00(-16) &$<$1.55(-16) \\
%\ion{Kr}{4} & 5346 &             &             & 2.28(-15) & 3.05(-15) &             &             \\
%\ion{Kr}{4} & 5868 &             &             & 3.48(-15) & 4.51(-15) &             &             \\
\enddata
\tablenotetext{a}{In units of erg s$^{-1}$ cm$^{-2}$, numbers in parentheses are exponents, colons after values indicates uncertain detections.}
\tablenotetext{b}{Observed flux.}
\tablenotetext{c}{Extinction-corrected flux.}
\end{deluxetable*}

The KPNO echelle spectra of NGC~6543 were taken over the three nights
18-20 June 2005 UT. We used the UV camera on the 4m Mayall Telescope
Cassegrain echelle spectrograph with echelle grating 79-63 and cross
disperser 226-1 with two different setups, each giving 20 km s$^{-1}$
resolution. The blue setup, used for the first two nights, covered the
wavelength range $\lambda\lambda$3200--5300\AA.  It used the cross
disperser in second order with a CuSO$_4$ order separating filter.  We
then switched to a red setup covering the range
$\lambda\lambda$4750--9900\AA\ in first order of the cross-disperser
grating, using a GG495 order separating filter.  For calibration we
observed standard stars HR 4468, HR 4963, HR 5501 and HR 8634 from
\citet{Ha94}, measured through a 6 arcsec wide slit.  The nights with
the blue setup were photometric, while there were some clouds present
when we used the red setup.  We used the measured strengths of
emission lines in the overlapping sections of the red and blue spectra
to scale the fluxes for the red spectra to match those of the blue
spectrum.
    
As was done at LCO, spectra were taken in two positions symmetrically
placed on either side of the central star.  In this case the slit was
2 arcsec wide by 10 arcsec long and was centered 3 arcsec to the E and
then 3 arcsec to the W of the central star with a slit position angle
of 90$^o$, as shown in Figure~\ref{fig6}.  As with Tc 1 these slit
positions are well inside the outer regions of the nebula.  The
combined exposure times at each slit position for each grating setting
were of order 60 min.  We flux calibrated the NGC~6543 spectra and
then averaged together the two slit positions to get a final spectrum
interpolated for the line of sight to the central star in the same way
as was done with the LCO spectra.  We determine the absolute
spectrophotometry of our calibration to have an accuracy of better
than 7 percent.  In Figures~\ref{fig7}-\ref{fig9} we show portions of
the nebular spectra of the three PNe that were used in the analysis of
the emission lines and which show both the strong diagnostic forbidden
lines and the weaker recombination lines of \ion{O}{2}.

\subsection{Nebular Emission Line Intensities} \label{sec4.2}

The emission line fluxes have been measured from the final co-added
and averaged spectra using the IRAF \textit{splot} routine.  The
measurements were straightforward because few of the lines of interest
showed evidence of significant blending.  The resultant observed
intensities for He2-138, NGC~6543, and Tc 1 are given in
Table~\ref{tab4} for lines that can be used to obtain $T_e$, $n_e$,
and extinction along the lines of sight.  The observed intensities
have been corrected for extinction by taking the flux ratios of
multiple unblended Balmer and Paschen line pairs from the same upper
levels and determining the logarithmic extinction at H$\beta$,
c$_{H\beta}$, from the expression
\begin{equation} \label{eq3}
c_{H\beta} = \left[X_{H\beta}/(X_1-X_2)\right] \times
\log_{10}\left(A_1 F_2 \lambda_2/A_2 F_1 \lambda_1\right)\,,
\end{equation}
where $A_{1,2}$, $\lambda_{1,2}$, and $F_{1,2}$ are the spontaneous
emission coefficients, wavelengths, and observed fluxes for a specific
Balmer and Paschen line pair, and $X_{1,2,H\beta}$ are the galactic
extinction law values fitted by \citet{Ho83} at the wavelengths of the
lines and H$\beta$ respectively (assuming $R=3.2$).  Individual
emission line fluxes were then corrected using the relation
\begin{equation} \label{eq4}
F_c(\lambda)=10^{c_{H\beta}X(\lambda)/X_{H\beta}}
F(\lambda)\,,
\end{equation}
where $F_c$ is the corrected flux.  Taking the average of values
obtained from multiple line pairs for our lines of sight we derive
values of c$_{H\beta}$ = 0.56, 0.14, and 0.33 for He2-138, NGC~6543,
and Tc 1.  These values are in good agreement with those of
\citet{CKS92}, who obtained global values of 0.40, 0.12, and 0.28 for
the three PNe.  The extinction corrected fluxes, $F_c$, are listed in
column 4 of Table~\ref{tab4}, including the upper limits to fluxes of
undetected lines, which have been taken to be the 3$\sigma$ rms flux
of the noise of the neighboring continuum.

\subsection{Plasma Diagnostics and Emission Measures} \label{sec4.3}

\begin{deluxetable*}{lll}
\tablecaption{Atomic Data References \label{tab5}}
\tabletypesize{\scriptsize}
\tablewidth{6in}
\tablehead{
\colhead{Species} & 
\colhead{Transition Probabilities} &
\colhead{Collision Strengths}
}
\startdata
C I  & \citet{NR79}                     & \citet{PA76}     \\
            & \citet{FFS85}                    & \citet{TN75}     \\
            &                                  & \citet{JBK87}    \\
N I  & \citet{Ze82}                     & \citet{BB81}     \\
            & \citet{FFT04}                    & \citet{DMR76}    \\
N II & \citet{NR79}                     & \citet{St94}     \\
            & \citet{WFD96}                    & \citet{SSS69}    \\
            &                                  & \citet{LB94}     \\
O I  & \citet{BZ88}                     & \citet{BB81}     \\
            & \citet{Me83}                     & \citet{Be88}     \\
            &                                  & \citet{LN76}     \\
O II & \citet{Ze82}                     & \citet{Pr76}     \\
            & \citet{WFD96}                    & \citet{MB93}     \\
O III& \citet{NS81}                     & \citet{Ag83}     \\      
            & \citet{WFD96}                    & \citet{ABT82}    \\
            &                                  & \citet{BBK80}    \\
            &                                  & \citet{BBK81}    \\
            &                                  & \citet{LB94}     \\
P II & \citet{KS86}                     & \citet{Ta04}     \\
	    & \citet{MZ82b}                    & \citet{KC70}     \\
S II & \citet{MZ82a}                    & \citet{Ke96}     \\
            & \citet{Ke93}                     & \citet{Me82}     \\
            & \citet{VVF96}                    &                  \\
S III& \citet{MZ82b}                    & \citet{Me82}     \\
            & \citet{HSC95}                    &                  \\
            & \citet{LL93}                     &                  \\
Cl III & \citet{MZ82a}                    & \citet{BZ89}     \\ 
Ar III & \citet{MZ83}                     & \citet{JK90}     \\
            &                                  & \citet{KC70}     \\
Ar IV & \citet{MZ82a}                    & \citet{ZLB87}    \\
            & \citet{KS86}                     &                  \\
Fe II & \citet{NS88}     & \citet{ZP95}     \\
(159 levels)& \citet{Ga62}                     & \citet{BP96}     \\
            & \citet{Na95}                     & \citet{BK01}     \\
            & \citet{SSK04}                    &                  \\
            & \citet{BK01}                     &                  \\*
Ni II & \citet{NS82}    & \citet{Ba04}     \\
(76 levels) & \citet{Ku92}                     &                  
\enddata
\end{deluxetable*}

\begin{deluxetable*}{lrrr}
\tablecaption{Electron Temperatures and Densities \label{tab6}}
\tablewidth{5.25in}
%\tabletypesize{\footnotesize}
\tablehead{
\multicolumn{1}{c}{Diagnostic} & 
\multicolumn{1}{c}{~He2-138} &
\multicolumn{1}{c}{NGC~6543} &
\multicolumn{1}{c}{~~~~Tc 1}  
}  
\startdata
\multicolumn{4}{c}{Density (cm$^{-3}$)} \\
\textrm{[}\ion{N}{1}] \lam5198/\lam5200    &  7000$^{+\infty}_{-6000}$ &   900$^{+900}_{-500}$ &   400$^{600}_{-300}$ \\
\textrm{[}\ion{S}{2}] \lam6716/\lam6731    & 15000$^{}_{}$ &  5000$^{+17000}_{-3000}$ &  3000$^{+3800}_{-1200}$ \\
\textrm{[}\ion{O}{2}] \lam3726/\lam3729    &  7500$^{+9000}_{-3000}$ &  6000$^{+2400}_{-1600}$ &  2000$^{+800}_{-600}$ \\
\textrm{[}\ion{Cl}{3}] \lam5518/\lam5538   &  7500$^{+10000}_{-3000}$ &  5000$^{+2100}_{-1400}$ &  3000$^{+1800}_{-1100}$ \\
\textrm{[}\ion{Ar}{4}] \lam4711/\lam4740   &  \nodata      &  4500$^{+1100}_{-900}$ &  \nodata      \\
\multicolumn{4}{c}{} \\
\multicolumn{4}{c}{Temperature (K)} \\ 
\textrm{[}\ion{O}{1}] (\lam6300+\lam6364)/\lam5577            &  9000$^{+1200}_{-70}$ & \nodata       & 14000$^{+2300}_{-1400}$ \\
\textrm{[}\ion{S}{2}] (\lam6716+\lam6731)/(\lam4069+\lam4076) &  6000$^{}_{}$ &  9000$^{+7000}_{-5000}$ &  9000$^{+5000}_{-3000}$ \\
\textrm{[}\ion{O}{2}] (\lam3726+\lam3729)/(\lam7320+\lam7330) &  7000$^{+3000}_{-2000}$ & 12000$^{+3000}_{-2000}$ & 10500$^{+2700}_{-1700}$ \\
\textrm{[}\ion{N}{2}] (\lam6548+\lam6583)/\lam5755            &  6500$^{+600}_{-700}$ & 10300$^{+800}_{-700}$ &  8500$^{+600}_{-500}$ \\
\textrm{[}\ion{S}{3}] \lam9069/\lam6312                       &  6000$^{+500}_{-300}$ &  8500$^{+500}_{-400}$ &  9500$^{+700}_{-500}$ \\
\textrm{[}\ion{Ar}{3}] (\lam7136+\lam7751)/\lam5192           & \nodata       &  8000$^{+400}_{-300}$ &  9000$^{+800}_{-500}$ \\
\textrm{[}\ion{O}{3}] (\lam4959+\lam5007)/\lam4363            & \nodata       &  8200$^{+200}_{-200}$ &  9000$^{+500}_{-400}$ \\
\enddata
\end{deluxetable*}

\begin{figure}[t]
\plotone{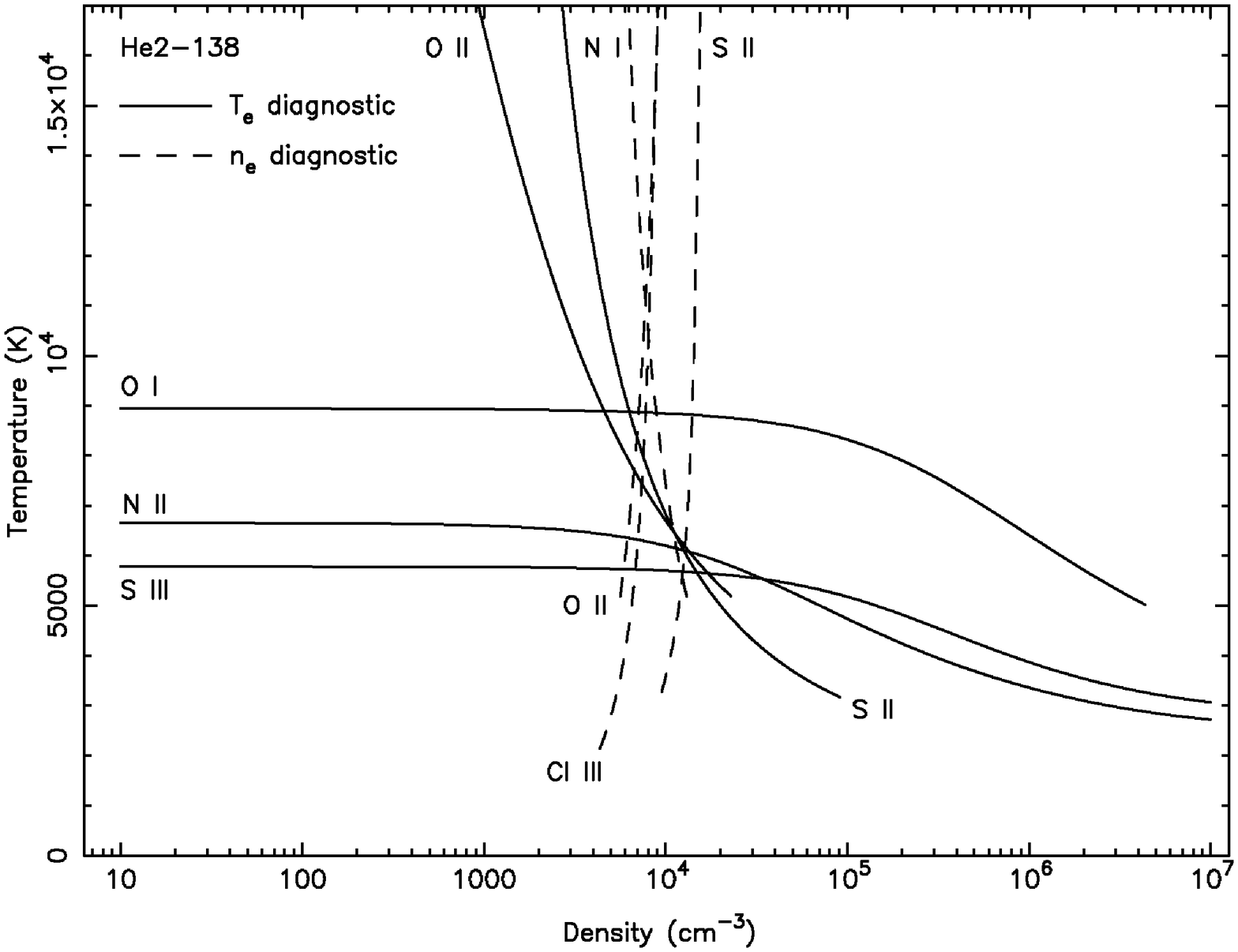}
\caption{Diagnostic diagram for He2-138.  Dashed lines indicate $n_e$
diagnostic curves and solid lines $T_e$ diagnostics curves derived
from emission line intensities listed in Table~\ref{tab4} input into
the ratios listed in Table~\ref{tab6}. \label{fig10}}
\end{figure}

\begin{figure}
\plotone{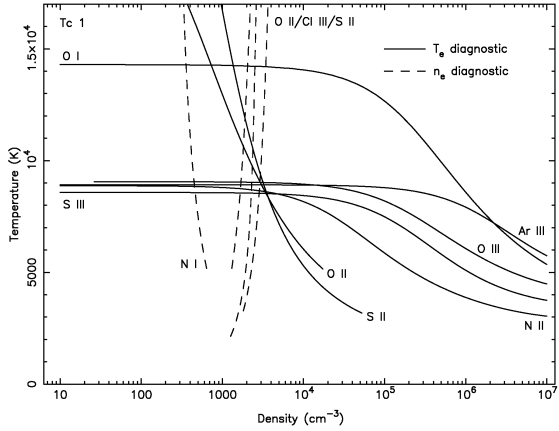}
\caption{Diagnostic diagram for Tc 1.  Dashed lines indicate $n_e$
diagnostic curves and solid lines $T_e$ diagnostics curves derived
from corresponding emission line intensities listed in
Table~\ref{tab4} input into the ratios listed in
Table~\ref{tab6}. \label{fig11}}
\end{figure}

\begin{figure}[t]
\plotone{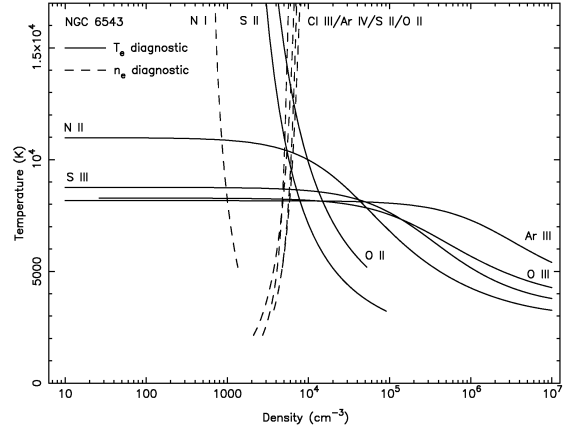}
\caption{Diagnostic diagram for NGC~6543.  Dashed lines indicate $n_e$
diagnostic curves and solid lines $T_e$ diagnostics curves derived
from corresponding emission line intensities listed in
Table~\ref{tab4} input into the ratios listed in
Table~\ref{tab6}. \label{fig12}}
\end{figure}

Emission measures and relative abundances of ions are normally
determined from their forbidden line intensities, which have a
sensitive dependence upon the kinetic temperature and density of the
gas.  Electron temperatures $T_e$ have been determined for regions of
different ionization primarily from the ratio of auroral to nebular
line intensities of [\ion{O}{1}], [\ion{S}{2}], [\ion{N}{2}],
[\ion{O}{2}], [\ion{O}{3}], [\ion{S}{3}], and [\ion{Ar}{3}] using
complete radiative and collisional multi-level calculations similar to
those in the IRAF \textit{nebular} package \citep{SD95}, as described
by \citet{Sh07} in their study of $s$-process elements in PNe.
Similarly, electron densities $n_e$ are sensitive to certain line
ratios such as [\ion{O}{2}] $\lambda$3726/29, [\ion{S}{2}]
$\lambda$6716/31, [\ion{Cl}{3}] $\lambda$5518/38, and [\ion{Ar}{4}]
$\lambda$4711/40.  We have used all of these forbidden line ratios,
when they were observed, and corresponding atomic data listed in
Table~\ref{tab5} to calculate appropriate values of $T_e$ and $n_e$ in
Figures~\ref{fig10}--\ref{fig12}.  The resulting values of $T_e$ and
$n_e$ are listed in Table~\ref{tab6} with their formal uncertainties.
The densities and temperatures derived from the different lines are
generally consistent with each other with the exception of the
[\ion{N}{1}] for NGC~6543 and Tc1, and the [\ion{S}{2}] density for
He2-138, as is evident from the plots in
Figures~\ref{fig10}--\ref{fig12}.  The disparate densities deduced for
He2-138 from the different lines could be real; the result of
inhomogeneities.  The [\ion{N}{1}] lines, on the other hand, are very
weak and thus the densities from that doublet are very uncertain.

In order to properly account for all relevant physical processes when
converting the observed emission line flux $F_c$ into the emission
measure we consider here the full definition of the emission measure.
The extinction-corrected flux of an optically thin emission line along
a line of sight is
\begin{equation} \label{eq5}
F_c = {\theta_o}^2 [h\nu_o/(4\pi)] \int n_u A_{ul}\, d\ell \,,
\end{equation}
where ${\theta_o}^2$ is the angular area of the gas being observed,
and $A_{ul}$ and $n_u$ are the line transition probability and number
density of the upper level.  The stronger forbidden transitions
normally have direct collisional excitation from the ground state as
the predominant mechanism exciting the line, therefore it is
convenient to write the equation of statistical equilibrium governing
the population of the upper level in terms of the ion and electron
densities as
\begin{equation} \label{eq6}
n_u \sum_{k<u} A_{uk}  =  n_e n_1 q_{1u}(T_e)  [1 +
  \xi(n_e,T_e,J_\nu) ] \,,
\end{equation}
where $q_{1u}(T_e)$ is the collision coefficient between the ground
state and upper level, and the term $\xi(n_e,T_e,J_\nu)$ represents
all other processes contributing to the population of the upper level,
e.g., radiative cascading and collisional population from upper
levels, collisional de-excitation to lower levels, continuum and
resonance fluorescence, and recombination.  We formulate the equation
this way in order to isolate the term $n_e n_1$, whose integral along
the line of sight is the emission measure of the ion.  For the
processes listed above, the expression for $\xi$ is
\begin{eqnarray} 
\xi(n_e,T_e,J_\nu) & = & \left[n_e n_1 q_{1u}\right]^{-1}
  \left\{\sum_{k>u} n_k A_{ku} + n_e \sum_{k>1,k\ne u}n_k q_{ku} \right. \nonumber \\ & & - n_e
  n_u \sum_{k\ne u} q_{uk} + \sum_{k\ne u} n_k J_{ku} B_{ku}
  \nonumber \\ & & \left. - n_u \sum_{k\ne u} J_{uk} B_{uk} +
  n_e n_{i+1} \alpha_u\right\}\,. \label{eq7}
\end{eqnarray}
Here, $J_{ku}$ and $B_{ku}$ are the mean intensity and Einstein $B$
coefficient for radiative (de-)excitation from level $k$ to level $u$,
and $\alpha_u$ is the effective recombination coefficient into level
$u$.

For most strong forbidden lines direct excitation from the ground
states predominates, and $\xi\ll 1$.  However, for certain ions, e.g.,
\ion{Fe}{2} and \ion{Ni}{2} \citep{Lu95,BPP96, BP96, Ba04}, and
certain transitions of CNO \citep{Gr75} other processes such as
radiative excitation and strong coupling between other excited levels
contribute to some of the stronger forbidden transitions, such that
$\xi>1$ for these lines under certain conditions.  The intensities of
these lines do not retain a simple linear dependence on $n_e n_1$, and
it is important to treat their excitation via detailed multi-level
calculations that involve the radiation field.

We have used detailed calculations of the relevant level populations
for all of the ions using the values of $T_e$, $n_e$, and $J_\nu$
appropriate for the level of ionization and augmented by incorporation
of additional levels and processes for the ions, to determine the
emission coefficients $q_{1u}(Te)$ and $\xi(n_e,T_e,J_\nu)$ for all of
the lines listed in Table~\ref{tab4}.  The radiation fields $J_\nu$ for
our slit positions have been taken from observations of the central
stars by IUE and FUSE for frequencies below the Lyman limit.  For
frequencies above the Lyman limit we have extrapolated the observed
stellar continua by assuming a black body flux at the appropriate
temperature for the central stars.  The dilution factors at our slit
positions were rather small so that radiative excitation by the
stellar continua was not competitive in the population of any level
that we considered.  For \ion{Fe}{2} and \ion{Ni}{2} we have used the
explicit multi-level population processes and cross sections of
Bautista, Pradhan, and collaborators to calculate line strengths for
these ions.  All known processes that might make significant
contributions to the line intensities have been included in the above
calculations, and corresponding values of $T_e$ and $n_e$ have been
used that are appropriate for the ionization state for each line.
Reddening corrected fluxes have then been used to compute ground state
emission measures $EM_i$ for those ions for which UV absorption was
also observed.

Combining equations~\ref{eq5}-\ref{eq7} produces the following
expression for the emission measure of an ion $i$,
\begin{eqnarray} \label{eq8}
EM_i  & \equiv  & \int n_e n_1 d\ell  \nonumber \\
      & = & 4\pi F_c \sum_{k<u} A_{uk} /
\left\{h\nu_o A_{ul} {\theta_o}^2  q_{1u}(T_e) \right. \nonumber \\
      & & \times \left.\left[1 +
    \xi(n_e,T_e,J_\nu)\right]\right\}\,.
\end{eqnarray}
The extinction-corrected fluxes from Table~\ref{tab4}, the atomic data
and cross sections from the references in Table~\ref{tab5}, and the
temperatures and densities listed in Table~\ref{tab6} appropriate to
the different ions depending upon their level of ionization have all
been used to determine the emission measures of ions from
equation~\ref{eq8} using the intensities of the various lines for our
sample of PNe.  The resulting emission measures are presented in
Table~\ref{tab7}.

\section{Comparative Column Densities from Emission and Absorption}
\label{sec5}

\subsection{Correction for Different Lines of Sight} \label{sec5.1}

\begin{deluxetable}{lcc}
\tablecaption{Forbidden Line Emission Measures \label{tab7}}
%\tabletypesize{\footnotesize}
\tablewidth{2.5in}
\tablehead{
\colhead{Species} &
\colhead{\lam (\AA)} &
\colhead{$EM_i$} \\
 & & (cm$^{-6}$ pc) 
}
\startdata \hline
\multicolumn{3}{c}{He2-138} \\ \hline
\ion{C}{1}  & 8727 & \phn0.404\phn    \\
\ion{P}{2}  & 7875 & $<$0.025\phn     \\
\ion{Fe}{2} & 5159 & \phn0.61\phn\phn    \\
\ion{Ni}{2} & 7378 & \phn0.0124       \\
\ion{O}{1}  & 6300 & 12.36\phn\phn \\
            & 6364 & 13.79\phn\phn    \\
\ion{S}{2}  & 4069 & 13.99\phn\phn    \\
            & 4076 & 14.92\phn\phn    \\
            & 6716 & 11.67\phn\phn \\
            & 6731 & 12.84\phn\phn \\
\ion{S}{3}  & 6312 & \phn3.93\phn\phn \\
            & 9069 & \phn3.89\phn\phn \\ \hline
\multicolumn{3}{c}{NGC~6543} \\ \hline
\ion{N}{1}  & 5198 & \phn0.029\phn    \\
            & 5200 & \phn0.031\phn    \\
\ion{O}{1}  & 6300 & \phn0.227\phn    \\
            & 6364 & \phn0.234\phn    \\
\ion{S}{2}  & 6716 & \phn0.202\phn    \\
            & 6731 & \phn0.201\phn    \\
\ion{S}{3}  & 6312 & \phn5.493\phn    \\
            & 9069 & \phn5.598\phn    \\ \hline
\multicolumn{3}{c}{Tc~1} \\ \hline
\ion{O}{1}  & 6300 & \phn0.060\phn    \\
            & 6364 & \phn0.073\phn    \\
\ion{S}{2}  & 4069 & \phn0.099\phn    \\
            & 4076 & \phn0.084\phn    \\
            & 6716 & \phn0.136\phn    \\
            & 6731 & \phn0.139\phn    \\
\ion{S}{3}  & 6312 & \phn0.736\phn    \\
            & 9069 & \phn0.990\phn       
\enddata
\end{deluxetable}

\begin{figure}[b]
\plotone{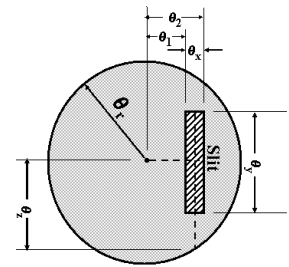}
\caption{A schematic indicating the relevant angles discussed in the
text that apply the derivations of the volume and shell correction
factors $\zeta_{\rm vol}$ and $\zeta_{\rm shell}$, assuming that the
nebula is a perfect sphere.}\label{fig13}
\end{figure}

In order to compare directly the results of the absorption and
emission abundance analyses the emission measures have to be converted
to effective column densities, or vice versa, by dividing the emission
measures by the electron density appropriate for each ion.  If we
designate $<\!\!n_e\!\!>_i$ as the mean value of $n_e$ in the emitting
region of the ion $i$, the equivalent emission column density for the
ion can be written as
\begin{equation} \label{eq9}
N_{em} = EM_i/(\zeta <\!\!n_e\!\!>_i)
\end{equation}
since the integrals that define the emission measure and the column
density of an ion differ only by the factor of the electron density in
the emission measure.  The constant $\zeta$ is a normalization factor
that corrects for the different path lengths along the two lines of
sight, viz., our emission line of sight passes through the entire
nebula whereas the absorption spectrum line of sight penetrates only
to the central star.  A derivation of $\zeta$ is given below that
allows for the fact that our observations record the flux within a
rectangle subtended by the entrance slit of the spectrograph, and this
slit is offset from the center of the nebula to avoid contamination of
the spectrum by the central star.

A generic representation of our emission-line measurement geometry is
shown in Figure~\ref{fig13}, with the idealization that the
appearance of the nebula is perfectly round in the sky.  We consider two
fundamentally different simplifications for the distribution of any
particular ion within the nebula.  The first such representation is a
uniformly filled sphere.  If the height of the slit $\theta_y$ is longer
than the chord through the nebula, the volume element $V$ of the sphere
that is interior to a projection of the slit with a width equal to
$\theta_x=\theta_2-\theta_1$ is given by
\begin{equation}\label{vol}
V={\pi\over 3}~\left[ \theta_1^3 - \theta_2^3 +
3\theta_r^2\left(\theta_2-\theta_1\right)\right]~.
\end{equation}
If we consider that the integration range in equation~\ref{eq5}
is over a distance equivalent to the angle subtended by $\theta_r$, we
can define the correction factor $\zeta_{\rm vol,0}$ to be simply $V$
derived above normalized to the volume within a rectangular box of
dimensions $\theta_x\theta_y\theta_r$; hence
\begin{equation}\label{zeta_full_vol}
\zeta_{\rm vol,0}={V\over\theta_x\theta_y\theta_r}={\pi \left[
\theta_1^3 - \theta_2^3 +
3\theta_r^2\left(\theta_2-\theta_1\right)\right]\over 3
\theta_x\theta_y\theta_r}~.
\end{equation}
If $\theta_y$ does not subtend the entire chord, we must reduce
$\zeta_{\rm vol,0}$ by a factor
\begin{equation}
F_{\rm vol}={1\over\pi}\left[ {\theta_y\over \theta_z}\left(
4-{\theta_y\over \theta_z^2}\right)^{\onehalf}
+2\sin^{-1}\left({\theta_y\over 2\theta_z}\right)\right]
\end{equation}
where
\begin{equation}
\theta_z=\left[\theta_r^2-\left({\theta_1+\theta_2\over
2}\right)^2\right]^{\onehalf}
\end{equation}
is the radius of the circle that represents the intersection of the
surface of the sphere with a plane that is aligned with the centerline
of the slit.  The factor $F_{\rm vol}$ is based on the approximation
that $\theta_x \ll \theta_z$, since it represents the area subtended by
lines bounded by $\theta_y$ inside the circle relative to the total area
of the circle, but only for a circle coincident with the slit
centerline.  The final value for $\zeta_{\rm vol}$ which applies to a
fully filled sphere is given by
\begin{equation}
\zeta_{\rm vol}=F_{\rm vol}\,\zeta_{\rm vol,0}
\end{equation}

Our second representation differs from the first in that the material is
assumed to be distributed in a thin shell, rather than throughout the
entire volume of the nebula.  In a development similar to the one that
we performed for the volume-filled nebula, we compute the area $A$ of a
projection of the slit on the surface of the sphere, under the condition
that it is long enough to cover the entire chord,
\begin{equation}
A=2\pi\theta_r\left(\theta_2-\theta_1\right)~.
\end{equation}
In this case, the normalization box has the dimensions
$\theta_x\theta_y$ multiplied by the thickness of the shell.  Since the
appropriate path length for equation~\ref{eq5}
in this case is the shell thickness, this thickness cancels out in the
equation for shell correction factor $\zeta_{\rm shell,0}$, leaving us
with the expression
\begin{equation}
\zeta_{\rm shell,0}={A\over
\theta_x\theta_y}={2\pi\theta_r\left(\theta_2-\theta_1\right)\over
\theta_x\theta_y}
\end{equation}
The reduction factor $F_{\rm shell}$ for $\theta_y < 2\theta_z$ is given
by
\begin{equation}
F_{\rm shell}={2\over\pi}\sin^{-1}\left({\theta_y\over 2\theta_z}\right)
\end{equation}
and this factor reverts to unity for $\theta_y \geq 2\theta_z$.
As before,
\begin{equation}
\zeta_{\rm shell}=F_{\rm shell}\,\zeta_{\rm shell,0}
\end{equation}

Values for the angles and correction factors $\zeta_{\rm vol}$ and
$\zeta_{\rm shell}$ are given in Table~\ref{tab8}, and are based upon
nebular diameters taken from the literature, as discussed below.  Both
values of $\zeta$ for NGC~6543 and Tc 1 are not very far from 2
because the slit heights were only about half the nebular diameters
and the slits were positioned rather close to the central stars.  The
relative change in going from $\zeta_{\rm vol}$ to $\zeta_{\rm shell}$
is large for He2-138 because the slit size was comparable to the
nebular diameter, and it was positioned near the nebula's edge.

The images of He2-138, NGC~6543, and Tc 1 in Figure~\ref{fig6} show
that all three PNe possess a degree of spherical symmetry for the
overall structure, but also have embedded asymmetrical, inhomogeneous
features, e.g., clumps and possible bipolar structure.  Differences
between the emission and absorption lines of sight therefore depend
not just on the footprint of the spectrograph slit on the nebulae and
whether the nebulae can be represented as filled shells or thin rings,
but also upon small-scale inhomogeneities that lie along one line of
sight but not the other.  Small-scale structure could well be the
dominant cause of differences between the separate lines of sight, and
such structure commonly depends on the level of ionization.

\begin{deluxetable}{
c    % quantity
c    % He2-138
c    % NGC~6543
c    % Tc1
}
\tablecolumns{4}
\tablewidth{0pt}
\tablecaption{Angles and 
Correction Factors \label{tab8}}
\tablehead{
\colhead{} & \multicolumn{3}{c}{Nebular Identification}\\
\colhead{Quantity\tablenotemark{a}} & \colhead{He2-138} & \colhead{NGC~6543} &
\colhead{Tc1}
}
\startdata
$\theta_r$&3.5&9.8&4.8\\
$\theta_1$&2.0&2.0&1.9\\
$\theta_2$&3.5\tablenotemark{b}&4.0&3.9\\
$\theta_y$&4.0&10.&4.0\\
$\zeta_{\rm vol}$&0.985&1.80&1.48\\
$\zeta_{\rm shell}$&4.12&2.22&2.64\\
\enddata
\tablenotetext{a}{Angles given in arc seconds.}
\tablenotetext{b}{One side of the slit extended beyond the edge of the
nebula ($\theta_2=4.0$), hence $\theta_2$ is set to $\theta_r$.}
\end{deluxetable}

\begin{deluxetable*}{lrccrc}
\tablecaption{Comparitive Column Densities from Emission and Absorption Lines \label{tab9}}
\tabletypesize{\small}
\tablewidth{6.5in}
\tablehead{
\multicolumn{1}{c}{Species} &
\multicolumn{1}{c}{$\log\,N_{abs}$} &
\multicolumn{1}{c}{$EM_i$} &
\multicolumn{1}{c}{$<n_e>_i$} &
\multicolumn{1}{c}{$\log\,N_{em}$ \tablenotemark{a}} &
\multicolumn{1}{c}{$\log\,N_{abs}-\log\,N_{em}$ \tablenotemark{a}} \\
 &
\multicolumn{1}{c}{(cm$^{-2}$)} &
\multicolumn{1}{c}{(cm$^{-6}$ pc)} &
\multicolumn{1}{c}{(cm$^{-3}$)} &
\multicolumn{1}{c}{(cm$^{-2}$)} &
}
\startdata \hline
\multicolumn{6}{c}{He2-138 ($\zeta_{\rm vol}=0.985$ $\zeta_{\rm shell}=4.12$)} \\ \hline
\ion{C}{1}  & 13.93$\pm0.16$    & \phn0.404\phn    &  \phn7000 & (14.26, 13.64) $\pm0.31$    & (-0.33, 0.29) $\pm0.35$ \\
\ion{P}{2}  & 13.58$\pm0.21$    & $<$0.025\phn\phn\phn &  \phn7000 & $<$(13.05, 12.43) $\pm0.42$ & $>$(0.53, 1.15) $\pm0.47$ \\
\ion{Fe}{2} & 14.37$\pm0.14$    & \phn0.61\phn\phn &  \phn7000 & (14.44, 13.82) $\pm0.64$    & (-0.07, 0.55) $\pm0.66$ \\
\ion{Ni}{2} & 12.77$\pm0.17$    & \phn0.0124   &  \phn7000 & (12.74, 12.12) $\pm0.21$    & (0.03, 0.65) $\pm0.27$ \\
\ion{O}{1}  & $>$15.48$\pm0.37$ & 12.8\phn\phn\phn      &  \phn7000 & (15.76, 15.14) $\pm0.34$    & $>$(-0.28, 0.34) $\pm$0.50       \\
\ion{S}{2}  & $>$15.49$\pm0.09$ & 12.3\phn\phn\phn      & 10000 & (15.58, 14.96) $\pm0.26$    &$>$(-0.09, 0.53) $\pm0.28$ \\
\ion{S}{3}  & 15.11$\pm$0.13    & \phn3.89\phn\phn     &  \phn7500 & (15.21, 14.59) $\pm0.11$    & (-0.10, 0.52) $\pm0.17$ \\ \hline
\multicolumn{6}{c}{NGC~6543 ($\zeta_{\rm vol}=1.80$ $\zeta_{\rm shell}=2.22$)} \\ \hline
\ion{O}{1}  & 13.48$\pm0.51$    & \phn0.23\phn\phn     &  \phn5000 & (13.89, 13.80) $\pm0.25$    & (-0.41, -0.32) $\pm0.57$ \\
\ion{N}{1}  & 12.29$\pm0.25$    & \phn0.030\phn    &  \phn4000 & (13.10, 13.01) $\pm0.21$    & (-0.81, -0.72) $\pm0.33$ \\
\ion{S}{2}  & 13.64$\pm0.07$    & \phn0.20\phn\phn     &  \phn5000 & (13.83, 13.74) $\pm0.14$    & (-0.19, -0.10) $\pm0.16$ \\
\ion{S}{3}  & 15.22$\pm0.14$    & \phn5.56\phn\phn     &  \phn5000 & (15.27, 15.18) $\pm0.04$    & (-0.05, 0.04) $\pm0.15$ \\ \hline
\multicolumn{6}{c}{Tc 1 ($\zeta_{\rm vol}=1.48$ $\zeta_{\rm shell}=2.64$)} \\ \hline
\ion{O}{1}  & 13.67$\pm0.06$    & \phn0.064\phn    &  \phn2000 & (13.82, 13.57)$\pm0.26$    & (-0.15, 0.10) $\pm0.27$ \\
\ion{S}{2}  & 13.67$\pm0.09$    & \phn0.137\phn    &  \phn3000 & (13.98, 13.73)$\pm0.14$    & (-0.31, -0.06) $\pm0.17$ \\
\ion{S}{3}  & 14.93$\pm0.07$    & \phn0.92\phn\phn     &  \phn3000 & (14.80, 14.55) $\pm0.04$    & (0.13, 0.38) $\pm0.08$ \\
\enddata
\tablenotetext{a}{Numbers in parentheses show the outcomes for
  $\zeta_{\rm vol}$ and $\zeta_{\rm shell}$, respectively.  These are
followed by the error limits arising from measurement uncertainties.}
\end{deluxetable*}

Our emission spectra provide information on the extent of both
large-scale geometrical and small-scale inhomogeneity effects from the
intensity variations of the emission lines along the slit length.  The
slit lengths we used were 4 and 10 arcsec with a spatial resolution of
1 arcsec along the slit, so we have relatively few independent spatial
resolution elements.  Nevertheless, the range in distance of the slit
from the central star over its length is comparable to the offset of
the slit center from the central star.  Thus, variations of intensity
along the slit due to the overall geometry of the nebulae should be
comparable to the intensity differences due to the different path
lengths of the absorption and emission sightlines.  We have measured
the variations in intensity of the [\ion{S}{3}] and [\ion{O}{1}]
lines, representing our highest and lowest ionization species, along
the slit in the three PNe.  We find for Tc 1 that both the
[\ion{S}{3}] and [\ion{O}{1}] lines have a very uniform distribution
of intensity throughout the full slit, and with no measurable
differences between the two slit positions.  Thus, for Tc 1 the
measurements indicate that the emission and absorption lines of sight
are likely to be very similar.

For He2-138 the [\ion{S}{3}] and [\ion{O}{1}] lines have virtually
identical, smooth intensity distributions where the intensity peaks at
the center and decreases outward toward the ends of the slit where it
falls off rapidly near the ends.  The intensity profile is more
characteristic of a filled volume than a thin shell distribution of
gas, but the smaller size of this nebula causes it to fill only 3
arcsec of the 4 arcsec slit length so the geometrical factor zeta
represents an important correction for this object.  Since there is no
indication of differences in the spatial distribution of the ions
based on their ionization level the geometrical normalization factor
zeta is the same for all of the lines.  The fact that the emission
spectra sampled the outer edge of the nebula makes the corresponding
geometrical correction for this PN rather uncertain, as was explained
in \S\ref{sec4.1.1}.  Thus, we consider the results for He2-138 to be
less reliable than those of Tc 1 and NGC~6543.
   
NGC~6543 presents a more complicated picture in terms of the
differences between the two lines of sight.  The [\ion{S}{3}]
completely fills the slits with a uniform intensity for one of the
slit positions, but shows variations of $\sim$25 percent in the other
position.  The [\ion{O}{1}] completely fills the slits also, but shows
large variations along the slit near the center.  Thus, the lowest
ionization species in this PN display a pronounced small-scale
structure that may cause the emission and absorption lines of sight to
be quite different for the lowest ionization lines.  Based on the
intensity variations one must admit the possibility of differences in
the column densities along different lines of sight for the neutral
species to be as large as factors of 3 for this object - an
uncertainty that compromises its usefulness for the neutral species
[\ion{O}{1}] and [\ion{N}{1}].

If the stellar absorption and nebular emission spectroscopy are
obtained at sufficiently high resolution one can use radial velocity
information from resolved line profiles to match velocity components
of optical emission with the corresponding UV absorption produced in
the same velocity intervals.  A comparison of these quantities within
the same velocity interval of the gas provides a more accurate
assessment of the comparative abundances than comparing the total
emission measure and column density integrated over the full profiles.
Local values of the emission column density can be determined for
specific kinematic regions within the nebulae.  Averaging these values
over the full velocity range for the nebular shell will produce the
global emission column density for the ion, and also information on
its fluctuations as a function of velocity.  Since thermal and
expansion velocities of nebulae are of order 10--20 km s$^{-1}$ a
spectral resolution less than 10 km s$^{-1}$ is optimal to perform the
analysis this way.  Our emission spectra lack the necessary spectral
resolution to perform such an analysis, and therefore we work with the
integrated (over wavelength) emission measures and column densities.

\subsection{Comparison of Absorption \& Emission Column Densities}
\label{sec5.2}

The absorption column densities obtained from UV resonance lines refer
specifically to those ions occupying the lower level of the
transition.  In order to obtain the total column density of the ion
the column densities for all the individual fine-structure levels of
the ground state must be summed together.  In our STIS spectra some
ions had blended absorption profiles for one or more of the
transitions from the ground state fine-structure levels that prevented
us from deriving the column densities for those levels.  We have
determined the column densities of ions in those levels by taking
values of $T_e$ and $n_e$ obtained from the emission lines for that
ion to solve for the level populations relative to the levels for
which column densities were determined.  Additionally, when more than
one resonance multiplet of an ion has yielded a column density we have
computed the mean value for the ion by weighting individual values
according to the inverse square of their uncertainties.  These
calculations, which have been applied to the absorption column
densities in Table~\ref{tab3}, yield the total ion column densities,
$N_{abs}$, to the central star, and these are listed in
Table~\ref{tab9} together with the formal errors that result from the
quantifiable uncertainties that are discussed below.

Emission measures for the same ions that have been observed in
absorption, and which appear in Table~\ref{tab7} for our sample of
PNe, are also presented in Table~\ref{tab9}.  For ions where more than
one forbidden line yields an emission measure we have determined the
average of the values, with stronger weight being given to the lines
of higher intensity and lower Boltzmann factor.  Using the
corresponding values of the electron density for each of the ions the
resulting emission column densities, $N_{em}$, have been determined
from the emission measures from equation~\ref{eq9}.  These are given
in the penultimate column of Table~\ref{tab9} together with the
combined errors, having been normalized to the absorption path length
by dividing by the geometrical factor $\zeta$.

A comparison of the values of $N_{abs}$ and $N_{em}$ for the different
ions from the two completely independent abundance analyses shows
moderately good agreement, with the exceptions of \ion{P}{2} in
He2-138 and \ion{N}{1} in NGC~ 6543.  Absolute abundances determined
from the forbidden emission lines and UV absorption lines give the
same results within $\pm$0.3 dex for adjacent lines of sight, which is
comparable to the combined formal errors of the analyses.  The
1$\sigma$ errors in the column densities derived from the analyses
represent the uncertainties that are quantifiable.  There are also
systematic errors that arise from assumptions rather than measurement
uncertainties, and both sources of error affect the accuracy of our
comparison of forbidden and recombination line column densities.

The primary sources of error for the absorption column densities are
(a) the determination of the proper continuum level, (b) the low S/N
of the intensities of weak absorption lines, (c) the insensitivity of
intensity to column density for saturated lines, and (d) the
determination of total ground state column density for states with
fine-structure levels when absorption from one or more of the levels
is either not observed or saturated.  The main sources of error for
the emission column densities are uncertainties in (a) the flux
calibration of the echelle spectra, which are at the 5--10 percent
level, (b) collision strengths for some of the forbidden lines, and
(c) the correct values of $T_e$ and $n_e$ that correspond to each of
the transitions, as assigned to the various ions from the diagnostics
shown in Figures~\ref{fig10}--\ref{fig12}.  The atomic data for most
of the forbidden lines that we have used for diagnostics and the
determination of column densities are believed to be known to better
than 30 percent accuracy.  With the exception of the [\ion{P}{2}] line
the current values for most of the forbidden line collision strengths
and transition probabilities are the result of calculations by
independent methods over the past three decades that have converged on
values that are in good agreement with each other and that have
changed little over the past five years.  Thus, the atomic data are
not likely to be major sources of error.  Rather, the largest sources
of formal errors in the emission column densities are uncertainties in
the values of temperature and density.  Because line intensities
depend upon these two parameters, errors in $T_e$ and $n_e$ translate
to errors in the column density.  Most of the lines are in the low
density limit and therefore the emission measures are rather
insensitive to density.  However, because of the Boltzmann factor the
line intensities are sensitive to $T_e$.  The errors caused by
uncertainties in the temperature, together with uncertainties in
intensity measurement and flux calibration, form the basis for the
combined error that is presented in Table~\ref{tab9} for each emission
column density.  To these uncertainties must be added the unknown
errors in collision strengths and those differences that small-scale
inhomogeneities may cause between the lines of sight.

Several features of Table~\ref{tab9} merit comment.  First, due to a
combination of weak, saturated, or strongly blended UV absorption
lines coupled with the failure of our nebular spectra to detect
forbidden lines from some ions, there are relatively few ions for
which we were able to derive independent abundances from both UV
absorption and forbidden emission lines.  Even with the relatively
long slit used to sample substantial portions of the PNe shells we
were not successful in detecting weak emission lines from a number of
the ions for which column densities had been measured from the STIS
spectra.

Second, with the exception of S$^{+2}$ all of the ion species listed
in Tables~\ref{tab7} and \ref{tab9} are the lowest ionization stages
that have ionization potentials greater than 13.6 eV.  This means that
some fraction of most of the ions we have measured could exist in
cold, neutral gas residing either within dense clumps embedded inside
the nebula or in a foreground shell of material around the nebula.
Such gas would complicate our analysis by increasing the absorption
column density without having any effect on the emission lines,
leading to legitimate differences between the emission and absorption
column densities.  Fortunately, we can test for this possibility by
comparing $N_{abs}$(\ion{O}{1}) with $N_{abs}$(\ion{S}{2}).  The
ionization fraction of O is closely coupled to that of H through a
charge exchange reaction that has a large rate constant
\citep{FS71,Ch80}, which guarantees that the amount of \ion{O}{1} in
the ionized gas is quite low and that practically all of the O is
neutral in \ion{H}{1} gas.  By contrast, in an H II region a
reasonable fraction of the S will be in the form of S$^+$ since its
ionization potential is high (23.3 eV).  Furthermore, in an \ion{H}{1}
region nearly all of the S should also be singly ionized. Therefore,
from \ion{H}{1} gas we expect to find the ratio
$N_{abs}$(\ion{O}{1})/$N_{abs}$(\ion{S}{2}) to be approximately equal
to the solar value of [O/S] = 1.46, assuming that neither of the
elements are significantly condensed onto dust grains nor are they
enriched or depleted by nuclear processes within the AGB progenitor of
the central star.  A ratio smaller than this value signifies
progressively less contribution to the column densities from neutral
gas.

For Tc 1 we have found that
$N_{abs}$(\ion{O}{1})$\approx$$N_{abs}$(\ion{S}{2}), which indicates
that any contribution from neutral material must be so small that it
can be neglected for our study.  The situation for NGC~6543 is not
quite so straightforward because our inferred value of
$N_{abs}$(\ion{O}{1}) for the nebula is based on the marginal
detection of \ion{O}{1}*.  Our ability to directly measure \ion{O}{1}
in the ground fine-structure level is compromised by possible
\ion{P}{2} $\lambda$1301.87 absorption from foreground ISM gas at a
velocity v = -14 km s$^{-1}$, which appears at the same wavelength as
the velocity-shifted \ion{O}{1} $\lambda$1302.17 line from the nebular
shell.  Nevertheless, we can derive an upper limit to the \ion{O}{1}
column density for the nebula from this feature, which has equivalent
width $EW = 10$ m\AA, by assuming the foreground \ion{P}{2} absorption
to be negligible.  When we do this, we derive a value $\log
N$(\ion{O}{1}) = 13.2, an amount that is above the lower bound for our
calculated $\log$ $N_{abs}$(\ion{O}{1}) that is listed in
Table~\ref{tab9}.  However, this value is still substantially lower
than our measurement of $N_{abs}$(\ion{S}{2}), so once again we are
assured that contamination of the column densities from neutral gas is
negligible for NGC~6543.

We are unable to make any assertion about
$N_{abs}$(\ion{O}{1})/$N_{abs}$(\ion{S}{2}) toward He2-138 because
both column densities were recorded as lower limits (the lines are
strongly saturated; see footnote c in Table~\ref{tab3}).  However, in
the spectrum of the central star for this nebula we see absorption
features from excited H$_2$ at v = -62 km s$^{-1}$, which is a strong
indication that we are viewing a photodissociation region at the inner
edge of a neutral shell surrounding the nebula.  Thus, it is possible
for this one object that $N_{abs}$ for ions that can exist within
\ion{H}{1} regions could add to the contributions from the ionized
nebula.  This may explain why
$N_{abs}$(\ion{P}{2})$\gg$$N_{em}$(\ion{P}{2}) for this PN, although
it is then puzzling why the discrepancies for \ion{Fe}{2} and
\ion{Ni}{2} are not nearly as large unless both of them are condensed
onto grains.

Finally, we point out that heavy element recombination lines for the
ion species that we studied by UV absorption, and which are
substantially weaker than the forbidden lines, remained under the
detection threshold of our spectra.  This limits our ability to make a
direct comparison of abundances determined from recombination lines
for our PNe.  However, although our spectra did not detect
recombination lines originating from any of the ions in
Tables~\ref{tab7} and \ref{tab9} from which UV resonance absorption
was observed, we did observe \ion{C}{2}, \ion{N}{2}, and \ion{O}{2}
recombination lines.  Any information that can be obtained from an
analysis of these recombination lines is potentially useful.  Of the
above ions only the \ion{O}{2} recombination lines originate from a
parent ion for which forbidden lines were observed, viz., O$^{+2}$, so
the column densities inferred from these lines are considered in the
following section.

\subsection{Recombination Line Column Densities} \label{sec5.3}

The results of Table~\ref{tab9} show that the absorption and forbidden
emission column densities agree within the uncertainties of
measurement error, inaccuracies in the values of temperature and
density, and inhomogeneities that cause the adjacent lines of sight to
sample different components of the nebulae.  This agreement indicates
that nebular analyses based upon forbidden emission lines yield heavy
element abundances that are the same as those derived from absorption
lines-the key result from this study.  That said, what conclusions can
be drawn from our sample of PNe about nebular abundances based on
recombination lines?  Do our objects show the same discrepancies
exhibited by other PNe?

\begin{deluxetable}{rcc}
\tablecaption{Recombination Line Abundances \label{tab10}}
\tabletypesize{\footnotesize}
\tablewidth{3.5in}
\tablehead{
\multicolumn{1}{c}{Line(Multiplet)} & 
\multicolumn{1}{c}{$F$\tablenotemark{a}} &
\colhead{ADF} \\
 & \multicolumn{1}{c}{(erg cm$^{-2}$ s$^{-1}$)} &
}
\startdata \hline
\multicolumn{3}{l}{NGC~6543} \\ \hline 
\ion{O}{2} $\lambda$4069.62,.88 (10) & 5.52(-14) & 2.24 \\
$\lambda$4072.15 (10)              & 4.32(-14) & 1.88 \\
$\lambda$4110.79 (20)              & 1.16(-14) & 5.01 \\
$\lambda$4153.30 (19)              & 2.72(-14) & 3.61 \\
$\lambda$4317.14 (2)               & 2.10(-14) & 2.59 \\
$\lambda$4345.56 (2)               & 2.47(-14) & 3.28 \\
$\lambda$4349.43 (2)               & 2.85(-14) & 1.57 \\
$\lambda$4638.86 (1)               & 2.91(-14) & 2.59 \\
$\lambda$4641.81 (1)               & 7.34(-14) & 2.87 \\
$\lambda$4649.14 (1)               & 8.59(-14) & 1.75 \\
$\lambda$4650.84 (1)               & 3.27(-14) & 3.21 \\
$\lambda$4661.63 (1)               & 3.20(-14) & 2.47 \\
Average                             &           & 2.8$\pm0.9$ \\ \hline
\multicolumn{3}{l}{Tc 1} \\ \hline 
\ion{O}{2} $\lambda$4069.62,.88 (10) & 4.75(-16) & 1.97 \\
$\lambda$4072.15 (10)              & 4.26(-16) & 1.82 \\
%$\lambda$4110.79 (V??)              & 1.16(-14) & 5.01 \\
$\lambda$4153.30 (19)              & 1.58(-16) & 2.12 \\
$\lambda$4317.14 (2)               & 2.11(-16) & 2.88 \\
$\lambda$4345.56 (2)               & 2.93(-16) & 4.09 \\
$\lambda$4349.43 (2)               & 2.98(-16) & 1.66 \\
%$\lambda$4638.86 (V1)               & 2.91(-14) & 2.59 \\
$\lambda$4641.81 (1)               & 7.08(-16) & 2.72 \\
%$\lambda$4649.14 (V1)               & 7.07(-16) & 1.36 \\
$\lambda$4650.84 (1)               & 3.73(-16) & 3.48 \\
$\lambda$4661.63 (1)               & 4.37(-16) & 3.18 \\
Average                             &           & 2.5$\pm0.9$ \\
\enddata
\tablenotetext{a}{Number in parentheses in an exponent.}
\end{deluxetable}

We have detected a number of the same \ion{O}{2} recombination lines
from NGC~6543 and Tc 1 that have been studied extensively in PNe over
the past decade and used to determine relative O$^{+2}$ abundances
\citep{Li00,RG05}.  Since the [\ion{O}{3}] forbidden lines are strong
in both objects it is straightforward to determine the abundance of
O$^{+2}$ as derived from the two types of lines.  No recombination
lines were observed in the spectrum of the very low ionization nebula
He2-138.

The permitted \ion{O}{2} lines from multiplets 1, 2, 10, 19, and 20
have been shown to be populated by recombination and to yield, among
themselves, consistent values of the O$^{+2}$ abundance in H II
regions and PNe \citep{Ts03,WLB05}.  We have used the
extinction-corrected intensities of these lines, the observed
intensities of which are shown in Table~\ref{tab10}, to compute the
O$^{+2}$ abundance relative to that determined from the [\ion{O}{3}]
$\lambda\lambda$5007, 4959 lines.  This ratio is referred to as the
``abundance discrepancy factor'' (ADF), and for PNe and H II regions
always has values that are greater than unity.  Using the same cross
sections and procedures described by \citet{RG05} and \citet{WLB05},
we have determined ADF values from the individual \ion{O}{2} lines,
and these are listed in the last column of Table~\ref{tab10}.  The
resulting mean values of the ADFs for NGC~6543 and Tc 1 are 2.8 and
2.5 (0.45 and 0.40 dex), respectively.  The mean ADF of 2.8 found here
for NGC~6543 is consistent with the previous determinations of ADF =
3.0 and 3.8 from other lines of sight through this same nebula
\citep{KLP96,WL04}.  Thus, at least two of our objects show the same
discrepancies between the recombination and forbidden line abundances
for O$^{+2}$ that are typical of PNe, and we have not by chance
studied nebulae for which the ADFs are close to unity.

Since there is good agreement between the absorption and forbidden
line column densities in our objects, one can infer from the above
results that recombination lines are likely to produce emission column
densities that are significantly higher than those derived from
absorption lines.  The final column of Table~\ref{tab9} shows that the
mean of the forbidden emission line column densities is marginally
larger than that of the absorption column densities for each of the
PNe.  The recombination line column densities would produce a larger
discrepancy.  However, because the ADFs for NGC~6543 and Tc 1 are of
the same size as the uncertainties in the column densities, a larger
sample of PNe is needed, especially including some objects with
relatively large ADFs, before a definitive statement can be made that
recombination abundances are not correct.  Given that we do not detect
any recombination lines from parent ions for which we measured UV
absorption lines, a direct comparison of absorption and recombination
line column densities for the same ions is likely to remain elusive.
Realistically, the only ions in Tables~\ref{tab7} and \ref{tab9} that
are likely to be parent ions of detectable recombination lines are
S$^+$ and S$^{+2}$.  With deeper spectra it should be possible to
observe the \ion{S}{2} recombination lines in our PNe, however the
relevant recombination coefficients are not known and are very
difficult to calculate with any accuracy (P. Storey, private
communication).

Emission line analysis of a large number of PNe has shown that
recombination lines originating from C$^{+2}$, N$^{+2}$, O$^{+2}$, and
Ne$^{+2}$ ions all yield roughly the same relative abundances among
themselves as do the collisionally excited forbidden and
intercombination lines from these same ions, and that the
recombination lines consistently indicate higher abundances with
respect to H and He \citep{RG05,WLB05,Li06}.  The discrepancies do not
appear to arise from problems with the atomic data, rather they seem
to be linked to characteristics that are specific to the nebulae.
Thus, the ADFs for doubly ionized CNONe tend to be approximately the
same in individual objects, and they vary in step with each other from
one nebula to the next although there are exceptions to this rule,
e.g., NGC~6720 \citep{GD01}.

If the agreement among the CNONe ADFs also applied to S$^{+2}$ one
could use the ADFs we have derived from the \ion{O}{2} lines in
NGC~6543 and Tc 1 to infer the recombination line column density for
S$^{+2}$, based on the [\ion{S}{3}] emission measure.  However, the
ADFs for elements in the 3rd row of the periodic table, including
sulfur, are virtually unknown because so few recombination lines are
detected and the relevant cross sections are not known \citep{Ba03}.
The only 3rd row ion for which a recombination abundance has been
determined is Mg$^{+2}$, from \ion{Mg}{2} lines having been measured
in ten PNe by \citet{Ba03}.  They found the Mg$^{+2}$/H$^+$ abundances
for their objects to show little evidence for enhancement over the
solar value, contrary to the large O$^{+2}$/H$^+$ enhancements derived
from the \ion{O}{2} recombination lines in the same PNe.  They
conclude that the recombination line abundance discrepancies may be a
phenomenon restricted to ions of the 2nd row of the periodic table,
viz., C, N, O, and Ne, and not exhibited by 3rd row ions.  

Our current study shows nonetheless that the electron densities and
temperatures deduced from the usual forbidden line analysis are
correct over a range of ionization zones that should include the
regions where C$^{+2}$, N$^{+2}$, O$^{+2}$, and Ne$^{+2}$ reside.
This means that the large ADFs for the second row elements cannot be
reflecting errors in the forbidden line abundances due to the use of
incorrect values of $n_e$ or $T_e$.  It would be necessary to find
another mechanism that would affect the forbidden line abundances
derived for C$^{+2}$, N$^{+2}$, O$^{+2}$, and Ne$^{+2}$, but not those
found for S$^+$ and S$^{+2}$.  In our opinion this makes factors
affecting the recombination line abundances almost certainly the cause
of the abundance discrepancies.

\section{Summary} \label{sec6}

The results reported here are derived from a limited sample of PNe and
are based upon observations that may not be extended in the near
future unless STIS is repaired and put back into service on HST.  For
this sample we find that the forbidden lines yield absolute abundances
for \ion{C}{1}, \ion{Fe}{2}, \ion{Ni}{2}, \ion{O}{1}, \ion{S}{2}, and
\ion{S}{3} that are consistent with those derived from their
absorption lines along adjacent sight lines.  Within the uncertainties
in the line intensities and calculations, the good agreement between
the column densities derived from the forbidden emission lines and the
UV absorption lines for the three PNe represents a validation of both
types of analysis.  It strengthens confidence in the abundances
derived from forbidden emission lines in spite of discrepancies with
the abundances derived from high level permitted recombination lines
from the same ions, and is the primary result of this investigation.

Although recombination lines were detected in only two of the three
objects in this study, those two PNe do show factor 2.5–-2.8
discrepancies between the O$^{+2}$ abundances derived from forbidden
lines and those from recombination lines.  This demonstrates that we
have studied PNe in which the abundance discrepancy problem exists.
Not being able to independently measure an abundance for O$^{+2}$ from
its UV resonance lines, which fall outside of the HST wavelength
range, we cannot confirm that the recombination abundances for
O$^{+2}$ are anomalously high.  Nor can we use the similarity in the
abundance discrepancy factors of C$^{+2}$, N$^{+2}$, and Ne$^{+2}$
with that of O$^{+2}$ to make a comparison of the inferred
recombination abundances of these ions with those from an absorption
analysis because their UV resonance lines also fall outside of the HST
wavelength range.  The one doubly ionized ion for which we do have
good absorption data, S$^{+2}$, did not have recombination lines
detected in our nebular spectra.

We \textit{have} shown that the electron densities and temperatures
deduced for a wide range of ionization levels do give correct
abundances using the forbidden lines from ions of other elements
within the same ionization zone as O$^{+2}$ (and C$^{+2}$, N$^{+2}$,
and Ne$^{+2}$).  In particular, the forbidden line abundances for
S$^{+2}$ are in good agreement with the absorption line abundances for
S$^{+2}$.  This is an important constraint since any explanation of
the ADF for O$^{+2}$ and other second row elements that implicates
errors in the forbidden line abundances would have to invoke a
mechanism that does not affect the forbidden line abundances of other
ions such as S$^{+2}$ in the same ionization zone.

RW, EBJ, and JAB gratefully acknowledge support for this research from
data reduction grant HST-GO-09736-A from STScI.  JAB and EP also
acknowledge financial support from NSF through grant AST-0305833.
Special thanks are due Jessica Kim and Paul Goudfrooij for their
skillful restitution of a STIS spectrum that was anomalous.

\end{document}